\documentclass[aps,prd,twocolumn,footinbib,reprint,superscriptaddress]{revtex4-2}
\usepackage[dvipsnames]{xcolor}
\usepackage{graphicx}
\usepackage{amsmath}
\usepackage{amsfonts}
\usepackage{dsfont}
\usepackage[german, english]{babel}
\usepackage{ulem}
\usepackage{color, xcolor}
\usepackage{float}
\usepackage{physics }
\usepackage[colorlinks=true,citecolor=blue,urlcolor=blue,linkcolor=blue]{hyperref}
\usepackage{bbold}
\usepackage{nccmath}
\usepackage{amssymb}

\newcommand{\secref}[1]{Sec.~\ref{#1}}
\newcommand{\appref}[1]{Appendix~\ref{#1}}
\newcommand{\figref}[1]{Fig.~\ref{#1}}
\renewcommand{\eqref}[1]{Eq.~(\ref{#1})}

\newcommand{\bk}{\boldsymbol{k}}
\newcommand{\bn}{\boldsymbol{n}}
\newcommand{\br}{\boldsymbol{r}}

\newcommand{\da}{^\dagger}
\renewcommand{\max}{\text{max}}
\newcommand{\tableref}[1]{Table~\ref{#1}}

\begin{document}

 \title{Multiple correlation lengths and type-1.5 superconductivity in $U(1)$ superconductors due to  hidden competition between irreducible representations of nonlocal pairing}
 \author{Anton Talkachov}
 \email{anton.talkachov@gmail.com}
 \affiliation{Department of Physics, KTH-Royal Institute of Technology, SE-10691, Stockholm, Sweden}
  \author{Paul Leask}
\affiliation{Department of Physics, KTH-Royal Institute of Technology, SE-10691, Stockholm, Sweden}
\author{Egor Babaev}
\affiliation{Department of Physics, KTH-Royal Institute of Technology, SE-10691, Stockholm, Sweden}
\affiliation{Wallenberg Initiative Materials Science for Sustainability, Department of Physics, KTH Royal Institute of Technology, SE-106 91 Stockholm, Sweden}
 \date{\today}

\begin{abstract}
A fundamental characteristic of a superconducting state is the coherence length $\xi$.
Multicomponent superconductors, particularly ones breaking multiple symmetries, are characterized by multiple coherence lengths. 
Here we show that even, nominally {\it single}-component superconductors under certain conditions are characterized by multiple coherence lengths.
We consider nearest-neighbor pairing interactions on a square lattice that leads to $s$-wave and $d$-wave representations of link superconducting order parameter.
We show that even if the subdominant order parameter is completely suppressed in the ground state, it results in multiple correlation lengths with nontrivial hierarchy, resulting in important physical consequences in inhomogeneous solutions.
Under certain conditions, this leads to type-1.5 superconductivity, where magnetic field penetration length falls between two coherence lengths, leading to vortex clustering in an external magnetic field.
\end{abstract}

\maketitle

The coherence length, introduced by Ginzburg and Landau \cite{GinzburgOld2} is a key characteristic of a superconducting state.
It is defined as an exponent that characterizes the spatial recovery of the Cooper pairing field away from a perturbation, such as a defect, local density fluctuation, or vortex.
In fact, this length scale is the key characteristic not only of a superconducting state but also of the normal state at temperatures close to the phase transition, where it sets the correlation length due to local pairing fluctuations and is important, e.g., in the phenomenon of paraconductivity \cite{larkin2005theory}.

Conventionally, superconductors were characterized by a single coherence (or, equivalently, correlation) length.
Then its ratio relative to another fundamental length scale, the magnetic field penetration depth $\lambda$, determines the magnetic response \cite{landau1950k,abrikosov1957magnetic,gor1959microscopic}.
This is the Ginzburg--Landau (GL) parameter $\kappa= \lambda/\xi$.
For $\kappa>1/\sqrt{2}$, one has a type-II superconductor that possesses stable vortices repelling each other and forming a vortex lattice.
In contrast, for $\kappa<1/\sqrt{2}$ we have a type-I superconductor where vortices attract each other and are not stable.
 
The clearest exception to this type-I/II dichotomy is where a superconductor has two broken symmetries, necessitating the introduction of two coherence lengths $\xi_1,\xi_2$.
For example, there are a plethora of superconductors that break $U(1)\times \mathbb{Z}_2$ time-reversal symmetry, giving rise to multiple coherence lengths (see detailed discussion in \cite{carlstrom2011length,Garaud.Corticelli.ea-2018a}). 
A new (termed in \cite{moshchalkov}, type-1.5) regime then arises when $\lambda$ is an intermediate length scale \cite{babaev2005semi}.
For a superconductor with $n$-broken symmetries the type-1.5 regime occurs when the magnetic penetration depth $\lambda$ falls between coherence lengths $\xi_i$ ($\xi_1 < \xi_2 < ... \lambda  ...< \xi_{n-1} < \xi_n$).
In such a state, vortices have a composite core with several competing length scales.
As a result of this competition, vortices attract each other on the length scale of larger coherence lengths, due to the vortex outer-cores overlapping.
On length scales of the magnetic penetration depth, the vortices repel one another due to magnetic/current-current interaction, forming stable vortex clusters.
This long-range attraction and short-range repulsion behavior, characteristic of type-1.5 superconductivity, enhances the superconducting typology by breaking the type-I/II dichotomy.
A more subtle case is the appearance of multiple correlation lengths in multiband superconductors that break a single $U(1)$ symmetry.
In this case, multiple well-defined correlation lengths are not guaranteed by symmetry, but instead require certain microscopic conditions to be met by interband interactions \cite{johan2,silaev1,silaev2,timoshuk2024microscopic}.
For experimental research see \cite{moshchalkov,moshchalkov2,ray2014muon,biswas2020coexistence,wang2025observation}.

In this work, we return to the basic question of what the coherence/correlation lengths and typology of a superconducting system with a single broken symmetry are.
We consider superconductivity in a system with competing pairing channels that belong to two different irreducible representations of nonlocal pairing.
In this case, we specifically focus on the regime where one of the competing order parameters is suppressed.
At first glance, such a system represents a case of a single-component superconductor that neither breaks multiple symmetries nor has multiple electronic bands.
Nevertheless, we show that, under certain conditions, a suppressed competing pairing channel significantly alters correlations in the system.
We show that such a system, in general, can only be described by the introduction of multiple correlation lengths.
The origin of additional coherence lengths in this type of ground state comes from the structure of gradient terms.

The effect is partially related to the phenomenon that, in a normal state, a proximate superconducting phase gives rise to well-defined coherence lengths, despite a lack of any broken symmetries.
In our case, we find that a proximate, but non-developed superconducting phase with different broken symmetry imprints additional correlation lengths in superconducting condensates.
We show that this principally alters the type of magnetic response of the system.
Furthermore, we show that such a nominally single-component system cannot be a type-II superconductor in proximity to a continuous phase transition to a different pairing state with additional broken symmetry.

We start from a single electronic band microscopic model with nearest-neighbor pairing and compute the phase diagram as a function of doping.
Two superconducting components arise in this model from the mathematical concept of symmetry (irreducible representations).
Then we derive the corresponding Ginzburg--Landau functional and discuss the coherence lengths $\xi_i$ and magnetic field penetration depth $\lambda$.
We estimate a region on the phase diagram where type-1.5 superconductivity arises.
Performing Ginzburg--Landau calculations, we show the existence of vortex clusters in a $U(1)\times \mathbb{Z}_2$ symmetry breaking state ($s+id$) and also report configurations with nonzero skyrmion charge.
Importantly, we find vortex clusters in the $U(1)$ symmetry breaking state where \textit{only one} component of the order parameter ($s$-wave or $d$-wave) is nucleated in the ground state.

\textit{Microscopic model.} Consider a mean-field Hubbard Hamiltonian on a square lattice with attractive interaction $V$ ($V>0$) between nearest neighbors:
\begin{equation}\label{eq:mean-field_Hamiltonian}
\begin{gathered}
    H_\text{MF} = -\sum_{i \sigma} \mu c_{i,\sigma}\da c_{i,\sigma}  -\sum_{\langle i,j \rangle, \sigma} e^{i\frac{ea}{\hbar}A_{ij}} c_{i,\sigma}\da c_{j,\sigma} \\
     + \sum_{\langle i,j \rangle} \left( \Delta_{ji} c_{i,\uparrow}\da c_{j,\downarrow}\da + \Delta^{*}_{ij} c_{i,\downarrow} c_{j,\uparrow} + \tfrac{|\Delta_{ij}|^2}{V}\right) + \sum_{\text{plaqu-} \atop \text{ettes}} \tfrac{a^2 L_z}{2 \mu_0 t_{xy}} B_z^2,
\end{gathered}
\end{equation}
where $\Delta_{ij} = -V \langle c_{i,\downarrow} c_{j,\uparrow} \rangle$ is the superconducting order parameter on a link $ij$.
Here $c_{i,\sigma}\da (c_{i,\sigma})$ is the creation (annihilation) operator for an electron with spin $\sigma$ on site $i$, $\mu$ is the chemical potential, $a$ is the unit cell size in $xy$ plane, $L_z$ is the interlayer distance, $\mu_0$ is the vacuum magnetic permeability, and $A_{ij}=\tfrac{1}{a} \int_j^i \boldsymbol{A}d\boldsymbol{l}$.
The energies, gap, chemical potential, and temperature are in the units of hopping integral $t_{xy}$, and $k_B=1$.

The superconducting order parameter living on nearest-neighbor links can be mapped to a linear superposition of two one-dimensional irreducible representations living on sites corresponding to extended $s$-wave ($A_{1g}$ irreducible representation) and $d$-wave ($B_{1g}$)
\begin{equation} \label{eq:gap}
    \Delta(\bk) = \Delta_{s} ( \cos k_x + \cos k_y ) + \Delta_{d} ( \cos k_x - \cos k_y ).
\end{equation}
Besides the overall $U(1)$ gauge symmetry, the superconducting gap is characterized by three parameters: the amplitudes $|\Delta_s|$, $|\Delta_d|$, and the relative phase difference between components $\theta = \arg (\Delta_s \Delta_d^*)$.
For a state  where $|\Delta_s| \neq 0, \, |\Delta_d| \neq 0$, the phase difference is $\pi/2$ for systems with the $C_4$ rotational symmetry \cite{ren1996ginzburg,musaelian1996mixed,shimahara2021stability}.

\begin{figure}[h]
    \begin{center}
    \includegraphics[width=0.99\columnwidth]{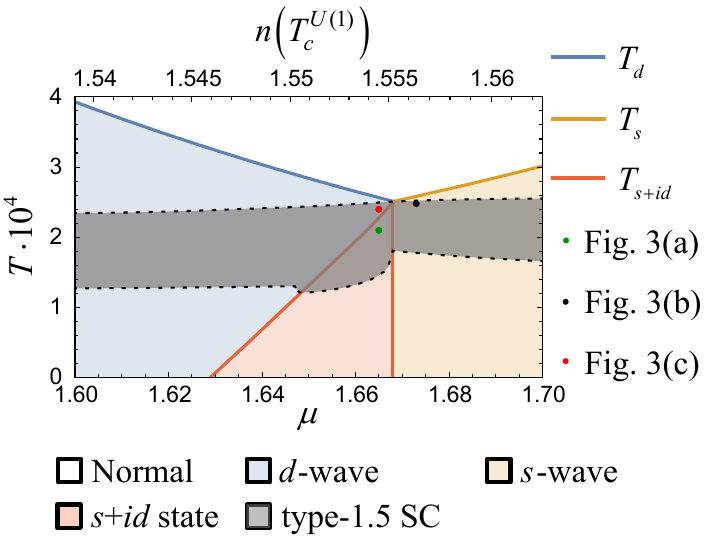}
    \caption{The phase diagram of the microscopic Hamiltonian \eqref{eq:mean-field_Hamiltonian} in chemical potential $\mu$ [or band filling at $U(1)$ critical temperature $n(T_c^{U(1)})$], temperature $T$ coordinates.
    The region for type-1.5 superconductivity is estimated based on simple approximations described in the text.
    The green, black, and red points correspond to ($\mu=1.665$, $T=2.1\cdot10^{-4}$), ($\mu=1.673$, $T=2.48\cdot10^{-4}$), and ($\mu=1.665$, $T=2.4\cdot10^{-4}$), respectively.
    In our microscopic calculations, the parameters we employed are: energy cutoff $\omega_D =0.1$, nearest-neighbor interaction strength $V=2$.
    Ginzburg--Landau calculation parameters: Unit cell size $a=0.4 \,$~nm, interlayer distance $L_z=1.3 \,$~nm, hopping parameter $t_{xy} = 0.01\,$~eV.}
    \label{fig:s+id phase diagram wD=0.1}
    \end{center}
\end{figure}
 
We have computed the superconducting phase diagram in band filling-temperature coordinates using numerical solutions of the self-consistency equation for the superconducting gap with small energy cutoff ($\omega_D = 0.1$, see \figref{fig:s+id phase diagram wD=0.1}).
A $d$-wave state is preferred near a half-filled band, $s$-wave for an almost filled band, and $s+id$ phase with relative phase difference $\pi/2$ in between.
A distinct feature is a sharp (vertical) transition between $s+id$ and $s$-wave phases, explained in \appref{app:microscopic description of phase diagram}.
Now, when we know the region of presence of $s$-wave and $d$-wave order parameters, we can investigate the system response to an external magnetic field.
We do this in a Ginzburg--Landau model that is derived in the vicinity of the $U(1)$ phase transition.

\textit{Ginzburg--Landau description.} The Ginzburg--Landau free energy density, derived from the microscopic model as an expansion in several small parameters associated with two gaps \eqref{eq:mean-field_Hamiltonian} in the weak coupling limit, has the following form
\begin{equation} \label{eq:GL free energy original}
\begin{gathered}
    F = \alpha_1 |\Delta_s|^2 + \alpha_2 |\Delta_d|^2 + \beta_1 |\Delta_s|^4 + \beta_2 |\Delta_d|^4 \\
    + \beta_3 |\Delta_s|^2 |\Delta_d|^2 + \beta_4 (\Delta_s^{*2} \Delta_d^2 + \Delta_s^2 \Delta_d^{*2}) \\
    + \gamma_1 |\boldsymbol{D} \Delta_s|^2 + \gamma_2 |\boldsymbol{D} \Delta_d|^2 \\
    + \gamma_{12}\left[ (D_y \Delta_s)^* (D_y \Delta_d) - (D_x \Delta_s)^* (D_x \Delta_d) + \text{c.c.} \right] \\
    + \tfrac{1}{2} (\nabla \times \boldsymbol{\tilde{A}})^2,
\end{gathered}
\end{equation}
where $D_i = \partial_i + iq \tilde{A}_i$ is a covariant derivative, $\boldsymbol{\tilde{A}} = \sqrt{\tfrac{L_z}{\mu_0 t_{xy}}} \boldsymbol{A}$, and $q=2 \frac{e a}{\hbar}\sqrt{\frac{\mu_0 t_{xy}}{L_z}}$.
The Ginzburg--Landau distance is in units of the unit cell size $a$.
Note that we expect the Ginzburg--Landau expansion to fail at low temperature.
Details of the microscopic derivation of the Ginzburg--Landau coefficients and characteristic length scales calculations below are presented in Appendices \ref{app:microscopic derivation} and \ref{app:coherence length}, respectively.

The resulting Ginzburg--Landau model is (i) anisotropic and (ii) there is an interband potential coupling ($\beta_{3,4}$) and mixed gradient coupling $\gamma_{12}$.
In general, these types of couplings create a hybridization of normal modes \cite{johan1,johan2,speight2021magnetic}: that is, a small variation in one of the gap fields causes a small variation in another, and vice versa.
Furthermore, in an anisotropic case, the resulting length scales should also depend on the chosen direction, i.e. fields recover their asymptotic form at different direction-dependent scales, away from a local perturbation \cite{winyard2019hierarchies}. 
Exact evaluation of length scales leads to a degree 6 polynomial equation (for amplitudes, phase difference, and three gauge field components) that determine all the length scales in the system, while all the normal modes are, in general, mixed in the models of the type of \eqref{eq:GL free energy original}  \cite{speight2021magnetic}.
For our purpose here, it is sufficient to make a rough estimate, that will follow by a numerical analysis: we first neglect the fact that the relative phase $\theta$ changes for small amplitude fluctuations and vice versa.
This assumption, in fact, breaks for vortex core analysis.
However, this method allows us to get a coarse estimate and to illustrate the principle.
For a precise description, we will resort to numerical calculations.
The magnetic field penetration depth $\lambda$ and coherence lengths $\xi_i$ are obtained by linearizing the free energy $F$ with respect to variations in the order parameter amplitudes ($|\Delta_{s/d}| = |\Delta_{s/d}^\text{GS}| + \sigma_{s/d}$ where $\sigma_{s/d}$ is small)
\begin{equation}
\begin{gathered}
    F_\text{lin} = F_\text{GS} + \frac{1}{2} \gamma_{ij}^{\alpha \beta} (D_i \sigma_\alpha)^*(D_j\sigma_\beta) + \frac{1}{2} \sigma_\alpha^* \mathcal{H}_{\alpha \beta} \sigma_\beta \\
    + q^2 |\boldsymbol{\tilde{A}}|^2 \left( \gamma_1 |\Delta_{s}^\text{GS}|^2 + \gamma_2 |\Delta_{d}^\text{GS}|^2 \right) + \frac{(\nabla \times \boldsymbol{\tilde{A}})^2}{2}
\end{gathered}
\end{equation}
where $\mathcal{H}$ is the Hessian matrix of potential energy about the ground state. 

The magnetic field penetration depth $\lambda$ in SI units can be \textit{roughly estimated} as
\begin{equation}
    \lambda^{-1} = 2 \frac{e}{\hbar}\sqrt{\frac{\mu_0 t_{xy}}{L_z}} \sqrt{2 \left(\gamma_1 |\Delta_s^\text{GS}|^2 + \gamma_2 |\Delta_d^\text{GS}|^2 \right)},
\end{equation}
Each amplitude eigenmode has a direction-dependent coherence length $\xi_i$ which, within the approximation, are given by
\begin{equation}
    \xi_i(\bn) = \frac{a}{\sqrt{\eta_i(\varphi)}},
\end{equation}
where $\varphi$ is the angle between the chosen direction $\bn$ and the $x$ axis and $\eta_i(\varphi)$ are eigenvalues of $\Gamma^{-1}(\varphi) \mathcal{H}$.
Here the matrix $\Gamma(\varphi)$ is given by
\begin{equation} \label{eq:gamma matrix}
    \Gamma(\varphi) = 2
    \begin{pmatrix}
        \gamma_1 && -\gamma_{12} \cos 2\varphi \\
        -\gamma_{12} \cos 2\varphi && \gamma_2
    \end{pmatrix}
\end{equation}
and describes the gradient energy for an amplitude perturbation in the direction $\bn$.
Note that off-diagonal terms describing mixed gradient terms vanish for diagonal propagation directions.
In general, the coherence lengths $\xi_i(\bn)$ are associated with linear combinations of $\sigma_s$ and $\sigma_d$, which gives rise to a hybridization of eigenmodes.
The presence of off-diagonal terms in $\Gamma(\varphi)$ is crucial: It ensures that the system has two coherence lengths for amplitudes even for a ground state with pure $s$-wave or pure $d$-wave, and suppressed subdominant component.

 \begin{figure}[t]
 \begin{center}
 \includegraphics[width=0.99\columnwidth]{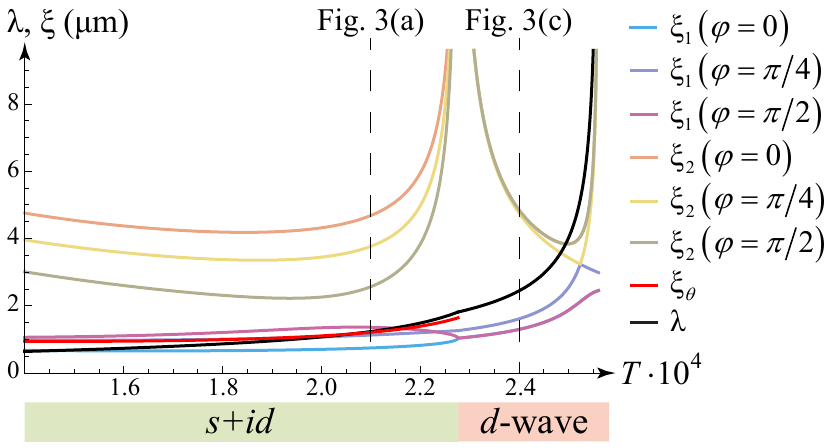}
 \caption{Temperature dependence of estimates of coherence lengths $\xi_{1(2)} (\varphi)$ for amplitude fluctuation in different directions defined by the angle $\varphi$ from the $x$ axis, coherence length $\xi_\theta$ for phase difference and magnetic field penetration length $\lambda$.
 Coherence length $\xi_\theta$ and magnetic field penetration length $\lambda$ are independent on the direction in the assumption of no mode mixing.  A coherence length for amplitude fluctuation diverges at superconducting phase transition, another divergent length scale occurs at time-reversal symmetry breaking transition. 
 This divergence occurs at both sides of the phase transitions, even when subdominant order parameter is suppressed.
 Type-1.5 superconductivity regime arises when magnetic field penetration length is the intermediate length scale, i.e.  when $\lambda$ (black line) is in between $\xi$'s (colored lines).
 Chemical potential $\mu=1.665$, other parameters are identical to \figref{fig:s+id phase diagram wD=0.1}.
 Calculations of coherence length for phase difference variations $\xi_\theta$ are presented in \appref{app:coherence length}.}
 \label{fig:xi and lambda}
 \end{center}
 \end{figure}

\begin{figure*}[t]
    \centering
    \includegraphics[width=0.95\textwidth]{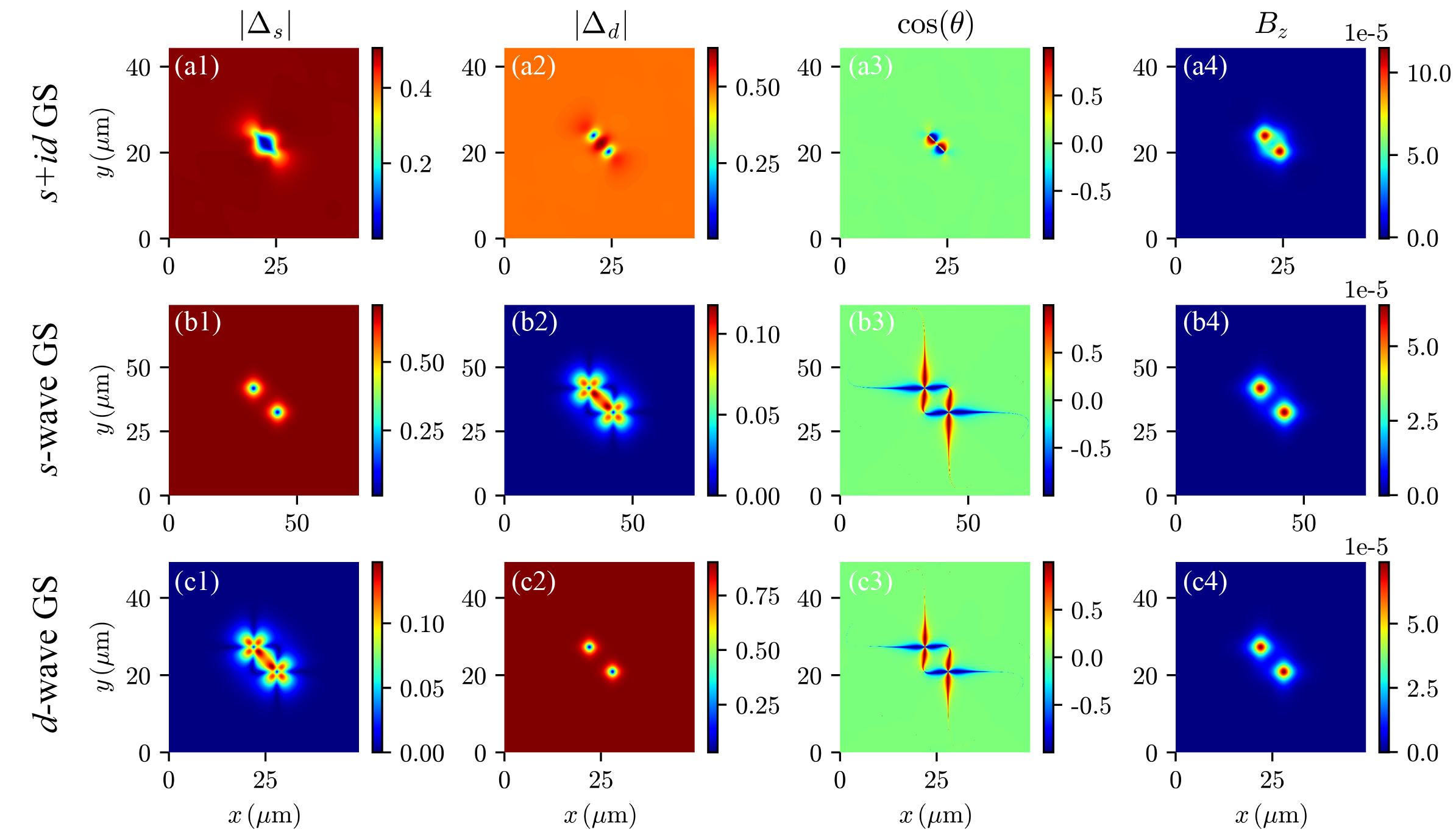}
    \caption{Two flux quanta vortex cluster solutions in Ginzburg--Landau model for three distinct ground states: (a) $s+id$ state, (b) pure $s$-wave, and (c) pure $d$-wave.
    The ground states (GSs) correspond to green, black, and red points in \figref{fig:s+id phase diagram wD=0.1}.
    Panels 1-4 show amplitudes $|\Delta_s|$, $|\Delta_d|$, relative phase difference $\theta=\arg (\Delta_s \Delta_d^*)$, and magnetic field $B_z$.
    Vortex interaction energy is negative that can be seen in \figref{fig: Binding energies}.
    The total energy density qualitatively follows the magnetic field shown in panels 4.
    Boundary conditions correspond to no current flowing through the
boundary, and the external magnetic field is zero.
    Gap amplitudes are presented in units $t_{xy} \sqrt{|\alpha_1|/\beta_1}$, magnetic field $B_z$ in $2 \hbar |\alpha_1|/(e a^2 \gamma_1)$.
    Model parameters are presented in \tableref{tab:model parameters}.}
    \label{fig: type-1.5}
\end{figure*}

Let us now compute the length scales in SI units for the unit cell size $a=0.4 \,$~nm, interlayer distance $L_z=1.3 \,$~nm, and hopping parameter $t_{xy} = 0.01\,$~eV.
These are typical values used for the 122-family of iron-based superconductors \cite{Rotter2008superconductivity} (as an example).
Figure \ref{fig:xi and lambda} shows the temperature dependence of the coherence lengths and magnetic field penetration depth for fixed chemical potential $\mu=1.665$ (vertical cross section through red and green points in \figref{fig:s+id phase diagram wD=0.1}).
The type-1.5 superconductivity regime arises when the magnetic penetration depth $\lambda$ (black line) is between coherence lengths $\xi$'s (colored lines).
Based on this constraint, we can calculate bounds for type-1.5 superconductivity and show these on the phase diagram in \figref{fig:s+id phase diagram wD=0.1}.
This method serves as an approximation, since our calculation of the length scale was approximate and does not prove type-1.5 behavior.
However, it provides an estimate for parameter space regions where exact calculation can be performed.
From Figs.~\ref{fig:s+id phase diagram wD=0.1} and \ref{fig:xi and lambda}, there is an indication that a type-1.5 regime could exist for $s+id$, pure $s$-wave, and pure $d$-wave ground states.
The latter two cases are the main focus of this letter.
One of the coherence lengths $\xi$ diverges at $U(1)$ and $U(1)\to U(1) \times \mathbb{Z}_2$ phase transitions, whereas the magnetic field penetration depth diverges only at the $U(1)$ phase transition.
Hence, type-1.5 superconductivity behavior is a generic type of superconductivity of double-transition superconductors.
However, there can be purely type-I behavior.

Since our estimates relied on analytical approximations, we next study vortex interaction behavior by performing numerical calculations in a microscopically derived Ginzburg--Landau model.
We pick three points in the phase diagram (\figref{fig:s+id phase diagram wD=0.1}) with $s+id$, pure $s$-wave, and pure $d$-wave ground states inside the \textit{estimated} type-1.5 regime.
The Ginzburg--Landau calculations presented in \figref{fig: type-1.5} show that vortices
form clusters.
The numerically calculated binding, or interaction, energy is negative for the three phase diagram points.
Hence, the type-1.5 regime is confirmed for all three points, which we estimated to display this hierarchy of the length scales.

The interesting aspect of the vortex clustering in a $s+id$ superconductor is its nontrivial morphology, that can be used to diagnose the symmetry of the pairing state and map phase diagram (for example of experimental morphology analysis see \cite{wang2025observation}).
Two, three and five vortex clusters form chains along diagonals for a minimal energy state (see additional plots for $N>2$ vortex configurations in \appref{app:additional data}).
The four vortex cluster is a diagonal chain for an $s+id$ ground state parameter set, whereas it forms a diamond for an $s$-wave ground state.
Diamond and chain vortex configurations have identical (up to numerical accuracy) energy for $d$-wave ground state with four vortices.
Hence, the preferred multi-vortex configurations are very sensitive to the model parameters.
On the contrary, $N=3, \, 4$ vortex clusters are regular polygons for isotropic multiband models.
Note that vortices exhibit type-1.5 behavior along any direction $\bn$ for the presented parameter sets.
However, vortex clusters aligned diagonally relative to crystalline axes have lower interaction energy.

\begin{figure}
    \centering
    \includegraphics[width=0.99\columnwidth]{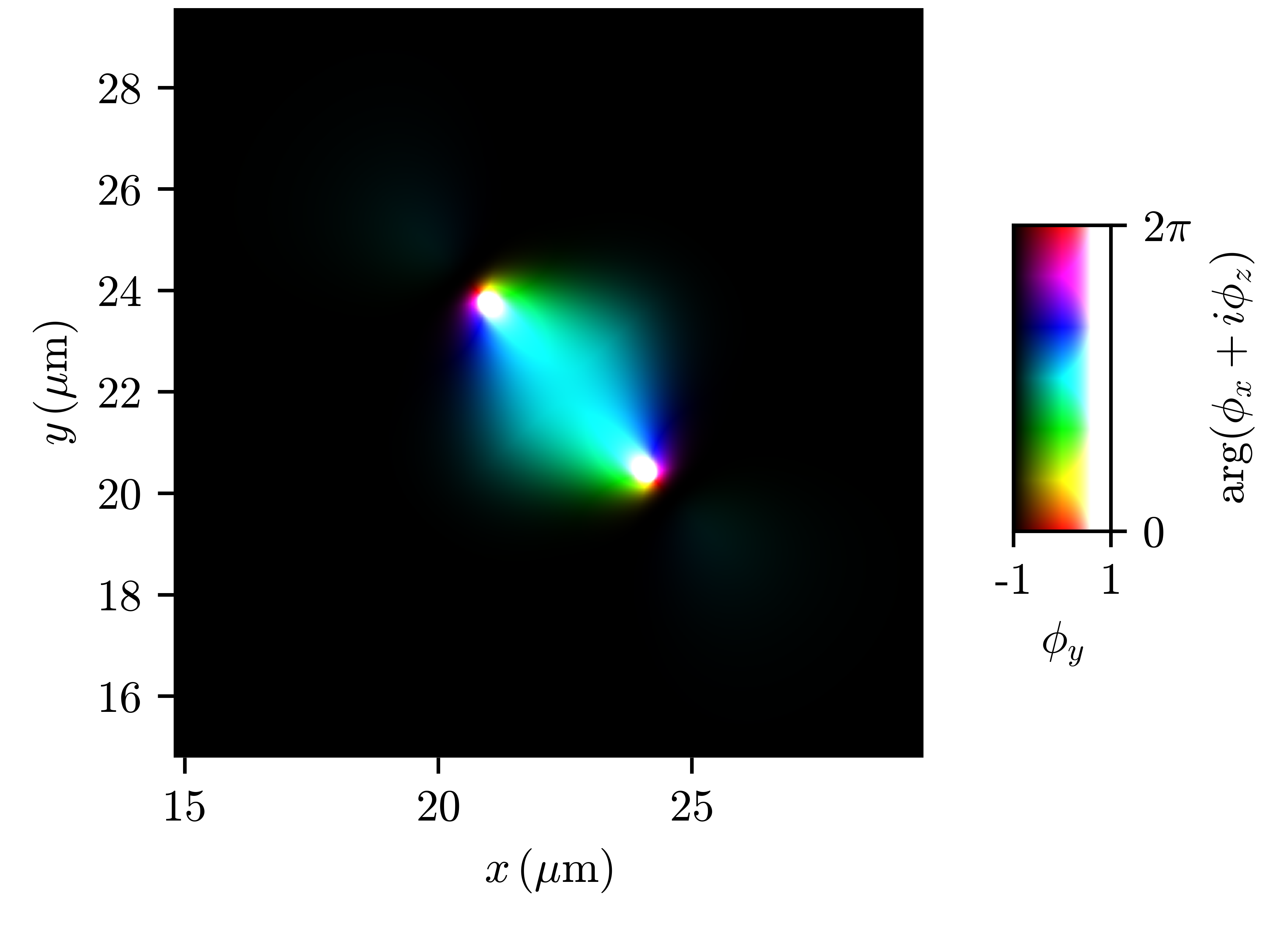}
    \caption{Plot of the pseudo-spin texture
    $\vec{\phi} = \frac{1}{|\Delta_s|^2+|\Delta_d|^2}\left(
        \Delta_s^*\Delta_d + \Delta_s\Delta_d^*, \,
        i(\Delta_s^*\Delta_d - \Delta_s\Delta_d^*), \,
        |\Delta_d|^2-|\Delta_s|^2      
    \right)$
    in the $s+id$ regime [\figref{fig: type-1.5}(a)].
    It shows that  the double-quanta   vortex is a skyrmion with topological charge $\mathcal{Q}=2$. In the type-1.5 regime it is a bound state of two $\mathcal{Q}=1$ skyrmions. The skyrmions are (iso-)rotated relative to one another such that the interaction energy between them is minimized. The skyrmion coloring is the standard Runge coloring scheme \cite{Leask_2022}.}
    \label{fig: Skyrmion}
\end{figure}

While we focus on length scales hierarchy and interactions, it is also useful to note that the vortex in \figref{fig: type-1.5}(a) (with $s+id$ ground state) carries a nontrivial skyrmionic topological charge, e.g. corresponding to a skyrmion.
Vortex cores in different components are \textit{spatially separated}, which can be seen from Figs.~\ref{fig: type-1.5}(a1), \ref{fig: type-1.5}(a2), and the pseudo-spin texture of the skyrmion is displayed in \figref{fig: Skyrmion}.
Such vortex core fractionalization can be detected by Scanning Tunneling Microscopy and was recently reported in a different system \cite{zheng2024direct}.
The two-vortex state has topological (skyrmion) charge $\mathcal{Q}=2$ (up to numerical accuracy). For discussion of skyrmions in other regimes and other superconducting models with broken time reversal symmetry see \cite{zhang2020skyrmionic,Garaud.Carlstrom.ea:11, Garaud.Carlstrom.ea:13}.
The skyrmion is composed of two $\mathcal{Q}=1$ skyrmions with relative orientation corresponding to their attraction.
Skyrmions are not unique features of type-1.5 superconductivity.
However, in the type-1.5 regime there is an additional attractive density-density interaction for skyrmions, giving rise to skyrmion clustering, just as for vortices.

\textit{Conclusion.} In this letter, we studied correlation lengths and vortex physics in a system with $s$- and $d$-wave pairing.
The particularly interesting aspect, not discussed in previous studies, is arising of multiple correlation lengths for a state that not only has one broken symmetry, but also a single order parameter in the ground state, leading to an observable physical effect: Type-1.5 behavior.
Namely, we showed that a competing, although suppressed order parameter, belonging to a different gap irreducible representation of nonlocal pairing, can principally alter the magnetic response of a system.
The extra length scale arises from the suppressed subdominant order parameter, which only nucleates near the vortex core, but gives a nontrivial hierarchy of the length scales of the problem.

This situation that we discussed should be more generic than our particular microscopic example.
Multiple correlation lengths can also arise in systems that have a single phase transition but are parametrically close to different superconducting instabilities belonging to irreducible representations (e.g. $d+ig$ state).

These effects can also be utilized in applications.
Instability due to vortex formation is utilized in single-photon detectors.  
Vortex clustering alters instabilities of  superconducting state and can 
be utilized in vortex-based superconducting nanowire single-photon detectors. 
For example, the recent study Ref.~\cite{bauer2025type} proposed that
the type-1.5 hierarchy of the length scales can diminish dark counts in single photon detectors.

\begin{acknowledgments}

This work was supported by the Knut and Alice Wallenberg Foundation via the Wallenberg Center for Quantum Technology (WACQT) and by the Swedish Research Council Grant 2022-04763. 
P.L. acknowledges funding from the Olle Engkvists Stiftelse through the grant 226-0103 and the Roland Gustafssons Stiftelse for teoretisk fysik.
E.B. was supported by Olle Engkvists Stiftelse a project grant from Knut och Alice Wallenbergs Stiftelse,  and partially by the Wallenberg Initiative Materials Science for Sustainability (WISE) funded by the Knut and Alice Wallenberg Foundation.

\end{acknowledgments}

 \appendix

\newpage

\section{Microscopic description of phase diagram} \label{app:microscopic description of phase diagram}

Self-consistency equation for superconducting gap has form
 \begin{equation} \label{eq:self consistency 1 component}
 \Delta(\bk) = \frac{1}{N} \sum_{\bk'} V(\bk, \bk') \Delta(\bk') \frac{\tanh{\frac{E}{2T}}}{2E},
 \end{equation}
 with only spin-singlet pairing channel
 \begin{equation}
 \begin{split}
 V(\bk, \bk') = &V \left( \cos k_x + \cos k_y\right) \left( \cos k_x' + \cos k_y' \right) \\
 &+ V \left( \cos k_x - \cos k_y\right) \left( \cos k_x' - \cos k_y' \right),
 \end{split}
 \end{equation}
 where $N$ is the number of sites in the lattice, $E = \sqrt{\xi^2 + |\Delta|^2}$ is the quasiparticle excitation energy, $\xi$ is the dispersion relation [$\xi(\bk) = -2 (\cos k_x + \cos k_y) - \mu$].
 The sum is computed for energies $E$ smaller than the upper bound denoted $\omega_D$.

Self-consistency \eqref{eq:self consistency 1 component} can be separated to the following system
 \begin{equation} \label{eq:self consistency 1 component explicit using sum}
 \begin{cases}
 \Delta_s = \Delta_s \frac{1}{N} \sum_{\bk} V ( \cos k_x + \cos k_y )^2 \frac{\tanh{\frac{E(\bk)}{2T}}}{2E(\bk)}, \\
 \Delta_d = \Delta_d \frac{1}{N} \sum_{\bk} V ( \cos k_x - \cos k_y )^2 \frac{\tanh{\frac{E(\bk)}{2T}}}{2E(\bk)}.
 \end{cases}
 \end{equation}
 Here, $s$ and $d$-waves are still implicitly coupled through energy $E$.
 The stationary solution of \eqref{eq:self consistency 1 component explicit using sum} is obtained when the sums in the RHS of Eq.~(\ref{eq:self consistency 1 component explicit using sum}) are equal to unity or an order parameter component ($\Delta_s$, $\Delta_d$ or both) is 0.
 For numerical calculations, the sum is presented as an integral over the first Brillouin zone ($\frac{1}{N} \sum_{\bk} = \int_\text{BZ} \frac{d\bk}{S_\text{BZ}}$ taking into account the limits of $\omega_D$).
 However, for analytical calculations, we use constant density of states [$\rho(\xi, \varphi) \approx N_F$] approximation similar to the BCS theory \cite{bardeen1957theory} [$\frac{1}{N} \sum_{\bk} =  \int_0^{2 \pi} \frac{d \varphi}{2 \pi} \int_{-\omega_D}^{\omega_D} d\xi \rho(\xi, \varphi) \approx 2 \int_0^{2 \pi} \frac{d \varphi}{2 \pi} \int_{0}^{\omega_D} d\xi N_F$].
 Another simplification is that form factors $( \cos k_x \pm \cos k_y )^2$ can be approximated as $\gamma_s \equiv \langle (\cos k_x + \cos k_y)^2 \rangle_\text{FS}$ for $s$-wave and $\Gamma_d \cos^2 2 \varphi$ ($\Gamma_d/2 \equiv \langle (\cos k_x - \cos k_y)^2 \rangle_\text{FS} = \gamma_d$) for $d$-wave \cite{hu2012local,xiang2022d}.
 Here $\langle ... \rangle_\text{FS}$ means the average over the Fermi surface (FS).
 Using the above-mentioned assumptions, self-consistency \eqref{eq:self consistency 1 component explicit using sum} for $s+id$ state becomes
 \begin{widetext}
 \begin{equation} \label{eq:self-consistency general}
 \begin{cases}
 \Delta_s = \Delta_s \cdot V N_F \gamma_s \int_0^{2 \pi} \frac{d \varphi}{2 \pi} \int_{0}^{\omega_D} d\xi  \frac{\tanh{\frac{\sqrt{\xi^2 + \Delta_s^2 + \Delta_d^2 \cos^2 2\varphi}}{2T}}}{\sqrt{\xi^2 + \Delta_s^2 + \Delta_d^2 \cos^2 2\varphi}}, \\
 \Delta_d = \Delta_d \cdot V N_F \Gamma_d \int_0^{2 \pi} \frac{d \varphi}{2 \pi} \cos^2 2 \varphi \int_{0}^{\omega_D} d\xi \frac{\tanh{\frac{\sqrt{\xi^2 + \Delta_s^2 + \Delta_d^2 \cos^2 2\varphi}}{2T}}}{\sqrt{\xi^2 + \Delta_s^2 + \Delta_d^2 \cos^2 2\varphi}}.
 \end{cases}
 \end{equation}
 \end{widetext}

Further, the system of equations is analyzed in special cases of low temperature and close to $U(1)$ critical temperature for states corresponding to pure $s$-wave ($\Delta_d = 0$), pure $d$-wave  ($\Delta_s = 0$), and $s+id$ state.
This is done to analytically address the shape of $s+id$ dome in the phase diagram in \figref{fig:s+id phase diagram wD=0.1}.

 \subsection{Characteristic temperatures for pure $s$-wave and $d$-wave $U(1)$ transitions}

 If $s$-wave and $d$-wave order parameters \textit{would not feel (or influence) each other} [$\Delta_d=0$ in the first of Eqs.~(\ref{eq:self-consistency general}) and $\Delta_s=0$ in the second equation], \textit{characteristic} temperatures for normal metal to superconductor transition are defined from independent equations
 \begin{subequations}
 \begin{align}
 &V N_F \gamma_s \int_0^{2 \pi} \frac{d \varphi}{2 \pi} \int_{0}^{\omega_D} d\xi  \frac{\tanh{\frac{\xi}{2T_s}}}{\xi} = 1, \\
 &V N_F \Gamma_d \int_0^{2 \pi} \frac{d \varphi}{2 \pi} \cos^2 2 \varphi \int_{0}^{\omega_D} d\xi \frac{\tanh{\frac{\xi}{2T_d}}}{\xi} = 1.
 \end{align}
 \end{subequations}
 The integral is a standard one with the result $\ln \frac{2 e^\gamma \omega_D}{\pi T_c}$, where $\gamma$ is Euler's constant.
 So the characteristic temperatures have form \cite{xiang2022d}
 \begin{subequations} \label{eq:both critical temperatures}
 \begin{align} \label{eq:s wave critical temperature}
 T_s &= \frac{2 e^\gamma}{\pi} \omega_D \exp\left(-\frac{1}{V N_F \gamma_s} \right), \\
 T_d &= \frac{2 e^\gamma}{\pi}  \omega_D \exp\left(-\frac{1}{V N_F \Gamma_d / 2} \right). \label{eq:d wave critical temperature}
 \end{align}
 \end{subequations}

 Remember that the result holds only for a normal metal to pure $s$-wave (or pure $d$-wave) transition.
 Therefore, only $\max(T_s,T_d)$ determines $U(1)$ symmetry breaking transition and can be called superconducting critical temperature.
 However, the concept and notation of the two \textit{characteristic} temperatures that \textit{do not influence each other} will be useful later in the section.

 \subsection{Pure $s$-wave to $s+id$ state transition for all temperatures} \label{sec:s to s+id}

 Transition from $U(1)$ broken symmetry state to $U(1) \cross \mathbb{Z}_2$ state is of the second order in the model at the mean-field level \cite{musaelian1996mixed,shapoval1996phase,ren1996ginzburg,liu1997mixed,ghosh1999two,roising2022heat}.
 We can assume $\Delta_s \neq 0, \, \Delta_d \rightarrow 0$ for $s+id$ state to $s$-wave transition.
 Self-consistency \eqref{eq:self-consistency general} on the transition line becomes
 \begin{equation}
 \begin{cases}
 1 = V N_F \gamma_s \int_{0}^{\omega_D} d\xi \frac{\tanh{\frac{\sqrt{\xi^2 + \Delta_s^2}}{2T}}}{\sqrt{\xi^2 + \Delta_s^2}}, \\
 1 = V N_F \gamma_d \int_{0}^{\omega_D} d\xi \frac{\tanh{\frac{\sqrt{\xi^2 + \Delta_s^2}}{2T}}}{\sqrt{\xi^2 + \Delta_s^2}}.
 \end{cases}
 \end{equation}
 If the first equation is satisfied, we know $\Delta_s$ for given $T, \, \mu$.
 The second equation is satisfied when $\gamma_d = \gamma_s$ ($\gamma_d = \Gamma_d / 2$) in the BCS limit.
 This coincides with the conclusion of Ref.~\cite{musaelian1996mixed} which deals only with $T=0$ case.
 The form factor $\gamma = \langle (\cos k_x + \cos k_y)^2 \rangle_\text{FS}$ depends only on the chemical potential $\mu$.
 Hence, the transition line does not depend on the temperature (assuming $T < T_s$).
 Therefore, the transition is a vertical line ($\mu = \text{const}$) on the $(\mu,T)$ phase diagram (see numerical calculation results in Figs.~\ref{fig:s+id phase diagram wD=0.1}, \ref{fig:s+id phase diagram appendix} and Refs.~\cite{ren1996ginzburg,maiti2015collective}).
A phase diagram with a sharp boundary between $s$-wave phase and nematic phase with broken time-reversal symmetry was discussed in connection of iron-based superconductors, see e.g.  \cite{ghosh2025elastocaloric}.

 Note that experimentally, one investigates electron density $n$ -- temperature $T$ phase diagram because the number of electrons in the sample is fixed during the measurement.
 The relation between electron density and chemical potential is the following
 \begin{equation}
 n = \frac{1}{N} \sum_{\bk} \left[1 - \frac{\xi}{E} \tanh\left( \frac{E}{2T} \right) \right],
 \end{equation}
 where $\xi$ depends on $\mu$ and $E$ depends on $\mu$ and $T$ [through $\Delta(T)$], $n \in [0;2]$.
 This electron density is a measure of the asymmetry between the positive and negative excitations of $\xi$.
 However, the change of electron density for $0 \leq T \leq T_c$ is quite small.
 Numerically, the change is of the order of $10^{-3}$ for $|n(\mu,T=0) - n(\mu, T_c)|$ (which is much smaller than typical values of  $n$).
 Hence, $(n,T)$ phase diagram will look qualitatively similar to $(\mu,T)$ phase diagram (see Refs.~\cite{timirgazin2019phase,talkachov2025microscopic}).

 \subsection{Pure $d$-wave to $s+id$ state transition for $(T_d - T) \ll T_d$}
 In the subsection, we are interested in a region where $d$-wave just developed ($\Delta_d \rightarrow 0$), and we are approaching $s+id$ state from $d$-wave phase ($\Delta_s = 0$).
 The characteristic temperature for $s$-wave is lower than for $d$-wave in the phase diagram region.
 Superconducting gap for $d$-wave can be approximately written in the way \cite{musaelian1996mixed, xiang2022d}
 \begin{equation} \label{eq:d wave gap high T}
 \Delta_d (T) = 4 \pi \sqrt{\frac{2}{21 \zeta(3)}} T \sqrt{- \ln{\frac{T}{T_d}}} \approx 3.537 T_d \sqrt{1 - \frac{T}{T_d}},
 \end{equation}
 where $\zeta$ is the Riemann zeta function.
 It follows from the second of \eqref{eq:self-consistency general}.

 The first of \eqref{eq:self-consistency general} can be simplified in a similar manner to Ref.~\cite{xiang2022d} that derives \eqref{eq:d wave gap high T}.
 First, Taylor expand integrand up to the second order in $\Delta_d$.
 Then integrate over $\xi$ using $\int_0^\infty dx \left( \frac{\tanh{x}}{x^3} - \frac{1}{x^2 \cosh^2{x}}\right) = 7 \zeta(3) / \pi^2$ (assuming $\omega_d \gg \Delta_d$) and substitute \eqref{eq:s wave critical temperature}:
 \begin{equation} \label{eq:d to s+id s-wave self-consistency high T}
 \begin{gathered}
 1 = V N_F \gamma_s \int_0^{2 \pi} \frac{d \varphi}{2 \pi} \int_{0}^{\omega_D} d\xi \frac{\tanh{\frac{\sqrt{\xi^2 + \Delta_d^2 \cos^2 2\varphi}}{2T}}}{\sqrt{\xi^2 + \Delta_d^2 \cos^2 2\varphi}} \\
 \approx V N_F \gamma_s \int_0^{2 \pi} \frac{d \varphi}{2 \pi} \left[ \int_{0}^{\omega_D} d\xi
 \frac{\tanh{\frac{\xi}{2T}}}{\xi} \right.  \\
 \left. - \int_{0}^{\omega_D} d\xi \left( \frac{\tanh{\frac{\xi}{2T}}}{2 \xi^3} - \frac{1 - \tanh^2{\frac{\xi}{2T}}}{4 T \xi^2}\right) \Delta_d^2 \cos^2 2\varphi \right] \\
 = V N_F \gamma_s \left[ \ln{\frac{T_s}{T}} - \frac{7 \zeta(3)}{16 T^2 \pi^2}  \Delta_d^2 (T) \right] + 1.
 \end{gathered}
 \end{equation}
 This equation only holds at the transition between pure $d$-wave to $s+id$ state.
 Combining Eqs.~(\ref{eq:d wave gap high T}) and (\ref{eq:d to s+id s-wave self-consistency high T}) one gets
 \begin{equation} \label{eq:d to s+id high T approximation}
 T_{d \text{ to } s+id} = \frac{T_s^3}{T_d^2} = \frac{2 e^\gamma \omega_D}{\pi} \exp \left[-\frac{1}{V N_F} \left(\frac{3}{\gamma_s} - \frac{2}{\Gamma_d / 2} \right) \right].
 \end{equation}

 This equation also defines a point where all four phases coexist ($T = T_s = T_d = T_{d \text{ to } s+id}$, $\gamma_s = \gamma_d = \Gamma_d/2$).
 At this point, the superconducting gap in both channels tends to zero.
A similar tetracritical point attracted recent interest   \cite{yuan2021strain}.
The transition temperature [$T_{d \text{ to } s+id} = T_s^3T_d^{-2}$] coincides with the result from an analysis of corresponding Ginzburg--Landau (GL) equations \cite{ren1996ginzburg}.

 \subsection{Pure $d$-wave to $s+id$ state transition for $T \ll T_d, \, T_s$}
 Self-consistency \eqref{eq:self-consistency general} becomes
 \begin{equation} \label{eq:d to s+id small T self-consistency}
 \begin{cases}
 1 = V N_F \gamma_s \int_0^{2 \pi} \frac{d \varphi}{2 \pi} \int_{0}^{\omega_D} d\xi \frac{\tanh{\frac{\sqrt{\xi^2 + \Delta_d^2 \cos^2 2\varphi}}{2T}}}{\sqrt{\xi^2 + \Delta_d^2 \cos^2 2\varphi}}, \\
 1 = V N_F \Gamma_d \int_0^{2 \pi} \frac{d \varphi}{2 \pi} \cos^2 2 \varphi \int_{0}^{\omega_D} d\xi \frac{\tanh{\frac{\sqrt{\xi^2 + \Delta_d^2 \cos^2 2\varphi}}{2T}}}{\sqrt{\xi^2 + \Delta_d^2 \cos^2 2\varphi}}.
 \end{cases}
 \end{equation}

Let us work with the first equation of \eqref{eq:d to s+id small T self-consistency}.
 Add and subtract term $\frac{1}{\sqrt{\xi^2 + \Delta_d^2 \cos^2 2\varphi}}$ to the integral:
 \begin{widetext}
 \begin{equation}
 \begin{split}
 \frac{1}{V N_F \gamma_s} &= \int_0^{2 \pi} \frac{d \varphi}{2 \pi} \int_{0}^{\omega_D} d\xi \left( \frac{\tanh{\frac{\sqrt{\xi^2 + \Delta_d^2 \cos^2 2\varphi}}{2T}}}{\sqrt{\xi^2 + \Delta_d^2 \cos^2 2\varphi}} - \frac{1}{\sqrt{\xi^2 + \Delta_d^2 \cos^2 2\varphi}} + \frac{1}{\sqrt{\xi^2 + \Delta_d^2 \cos^2 2\varphi}} \right) \\
 &= \int_0^{2 \pi} \frac{d \varphi}{2 \pi} \int_{0}^{\omega_D} \frac{d\xi}{\sqrt{\xi^2 + \Delta_d^2 \cos^2 2\varphi}} \left(\tanh{\frac{\sqrt{\xi^2 + \Delta_d^2 \cos^2 2\varphi}}{2T}} -1\right) + \ln \frac{4 \omega_D}{\Delta_d}.
 \end{split}
 \end{equation}
 \end{widetext}
 where the second integral $\int_0^{2 \pi} \frac{d \varphi}{2 \pi} \int_{0}^{\omega_D} \frac{d \xi}{\sqrt{\xi^2 + \Delta_d^2 \cos^2 2\varphi}} = \int_0^{2 \pi} \frac{d \varphi}{2 \pi} \ln \frac{2 \omega_D}{\Delta_d |\cos 2\varphi|} = \ln \frac{4 \omega_D}{\Delta_d}$ is computed in the limit $\omega_D \gg \Delta_d(0)$ \cite{xiang2022d}.
 In the first integral, we change coordinates to $x = \beta \xi$ ($\beta = T^{-1}$) and $y = \beta \Delta_d \cos 2 \varphi$
 \begin{equation}
 \begin{split}
 \frac{1}{V N_F \gamma_s} = &\ln \frac{4 \omega_D}{\Delta_d} + \frac{1}{2\pi} 8 \int_0^{\beta \Delta_d} \frac{dy}{2 \sqrt{\beta^2 \Delta_d^2 - y^2}} \\
 & \cdot \int_0^{\beta \omega_D} \frac{dx}{\sqrt{x^2 + y^2}} \left(\tanh \frac{\sqrt{x^2 + y^2}}{2} - 1 \right).
 \end{split}
 \end{equation}
 The main contribution to the integral comes from the region where $x$ and $y$ are the smallest.
 Therefore, we can use approximation $\sqrt{\beta^2 \Delta_d^2 - y^2} \approx \beta \Delta_d$
 \begin{equation} \label{eq:small T intermediate step 1}
 \begin{split}
 \frac{1}{V N_F \gamma_s} = &\ln \frac{4 \omega_D}{\Delta_d} + \frac{2 T}{\pi \Delta_d} \int_0^{\beta \Delta_d} dy \\
 &\cdot \int_0^{\beta \omega_D} \frac{dx}{\sqrt{x^2 + y^2}} \left(\tanh \frac{\sqrt{x^2 + y^2}}{2} - 1 \right).
 \end{split}
 \end{equation}
 The upper bounds for both integrals can be approximated as infinite because $\beta \rightarrow \infty$ and $\Delta_d, \, \omega_D$ are some not infinitesimal parameters.
 The integral can be computed in the following way:
 \begin{equation}
 \begin{gathered}
 \int_0^{\infty} dy \int_0^{\infty} \frac{dx}{\sqrt{x^2 + y^2}} \left(\tanh \frac{\sqrt{x^2 + y^2}}{2} - 1 \right) \\
 = \int_0^{\pi / 2} d \varphi \int_0^\infty dr \left(\tanh \frac{r}{2} - 1 \right) = -\pi \ln 2.
 \end{gathered}
 \end{equation}
 Let us substitute this result to \eqref{eq:small T intermediate step 1} and substitute \eqref{eq:s wave critical temperature} to the LHS:
 \begin{equation}
 \ln \frac{2 e^\gamma \omega_D}{\pi T_s } = \ln \frac{4 \omega_D}{\Delta_d} - \frac{2 T \ln 2}{\Delta_d}
 \end{equation}

 Therefore, temperature for the pure $d$-wave to $s+id$ state transition is
 \begin{equation}
 T_{d \text{ to } s+id} = \frac{\Delta_d}{2 \ln 2} \ln \frac{2 \pi T_s}{\Delta_d e^\gamma}
 \end{equation}
 The last step is to substitute the superconducting gap at zero temperature ($\Delta_d (0) =
 2 \pi e^{-\gamma - 1/2} T_d$ \cite{xiang2022d, musaelian1996mixed}) instead of $\Delta_d$:
 \begin{equation} \label{eq:d to s+id transition low T}
 \begin{gathered}
 T_{d \text{ to } s+id} = \frac{\pi e^{-\gamma - 1/2} T_d}{\ln 2} \ln \frac{T_s e^{1/2}}{T_d} \\
 = \frac{2 \omega_D}{\sqrt{e} \ln 2} \left(\frac{1}{2} + \frac{1}{V N_F \Gamma_d / 2} - \frac{1}{V N_F \gamma_s} \right)  e^{-2/V N_F \Gamma_d}.
 \end{gathered}
 \end{equation}
 For zero temperature of the transition, we have 
  \begin{equation} \label{eq:d to s+id T=0}
 \frac{1}{2} =  \frac{1}{V N_F} \left( \frac{1}{\gamma_s}- \frac{2}{\Gamma_d} \right)  = \ln \frac{T_d}{T_s}.
 \end{equation}
 So the transition happens when $T_d = T_s e^{1/2}$ if one uses auxiliary characteristic temperature for $s$-wave (which does not correspond to a phase transition).
 The left part of \eqref{eq:d to s+id T=0} was also obtained in Ref.~\cite{musaelian1996mixed}.

 \subsection{Numerical calculations of a phase diagram and comparison to analytical results}
\begin{figure}[h]
 \begin{center}
 \includegraphics[width=0.99\columnwidth]{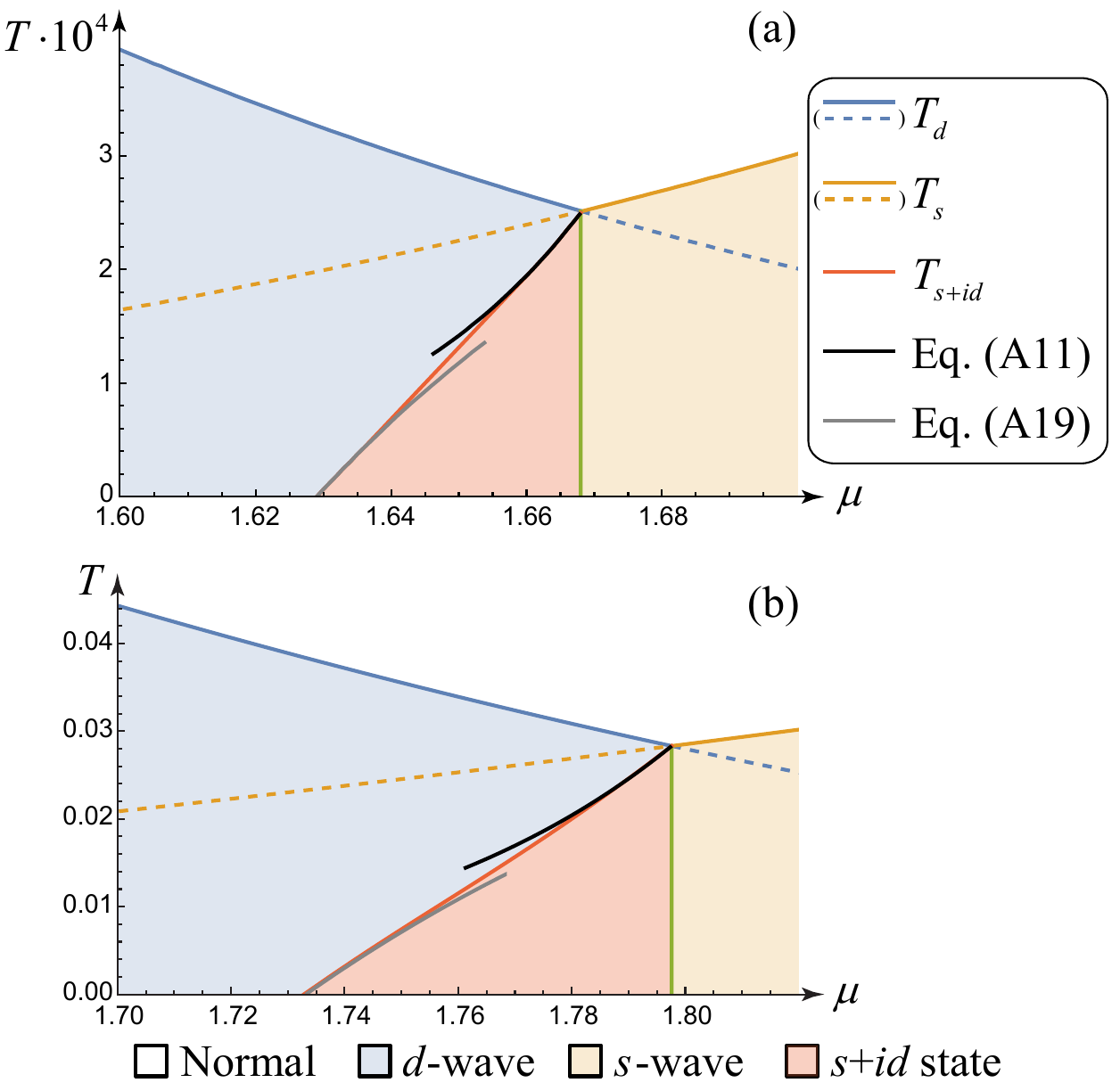}
 \caption{The phase diagram of the microscopic Hamiltonian \eqref{eq:mean-field_Hamiltonian} in chemical potential $\mu$, temperature $T$ coordinates.
 (a) energy cutoff $\omega_D =0.1$, (b) $\omega_D \rightarrow \infty$ (full Brillouin zone contributes).
 Orange and blue dashed lines for $T_s$ and $T_d$ do not correspond to phase transitions.
 Nearest-neighbor interaction strength $V$ is 2.}
 \label{fig:s+id phase diagram appendix}
 \end{center}
 \end{figure}

Let us compute superconducting phase diagram in $(\mu, T)$ coordinates using numerical solution of the Eqs.~(\ref{eq:self consistency 1 component explicit using sum}) with a small energy cutoff [$\omega_D = 0.1$, see \figref{fig:s+id phase diagram appendix}(a)].
 A distinct feature is a sharp (vertical) transition between $s+id$ and $s$-wave phases, predicted in \secref{sec:s to s+id}.
 Numerically, we see that $\mu_{s \text{ to } s+id}$ varies, but the chemical potential change between $T=0$ and $T=T_s=T_d$ is $O(10^{-5})$ and close to the computational precision.
 High and low temperature approximations [Eqs.~(\ref{eq:d to s+id high T approximation}) and (\ref{eq:d to s+id transition low T}), respectively] describe $d$-wave to $s+id$ state transition quite well.
 They hold far beyond the temperature range assumptions made for their derivation.

As a limiting type of check, we use the assumption of no energy cutoff (for a square lattice $\omega_D > 8$ has the same properties).
 This means that all excitations from the dispersion relation [$\xi(\bk) = -2 (\cos k_x + \cos k_y) - \mu$] contribute to the Cooper pairing [in the \eqref{eq:self consistency 1 component explicit using sum} all momenta are considered].
 The numerically obtained phase diagram in coordinates $(\mu,T)$ is presented in \figref{fig:s+id phase diagram appendix}(b).
 High temperature and low temperature approximations [Eqs.~(\ref{eq:d to s+id high T approximation}) and (\ref{eq:d to s+id transition low T}), respectively] are plotted in black and gray colors.
 They are in good agreement with numerical results even beyond the validity of the approximations.
 Namely, \eqref{eq:d to s+id transition low T} works for $T < 0.2 T_d$ ($T < 0.3 T_s$) compared to initial constraint $T \ll T_d, \, T_s$.
 And \eqref{eq:d to s+id high T approximation} works for $T > 0.8 T_d$ compared to $(T_d - T) \ll T_d$.
 Recall that analytical results were derived within the BCS approximation ($T \ll \omega_D \ll 1$).
 However, as we can see, the results can also be applied to large quasiparticle energy excitations using effective values of superconducting form factors.

 \section{Derivation of Ginzburg--Landau equations from microscopic model} \label{app:microscopic derivation}

The derivation of GL equations is based on expanding Gor’kov’s (microscopic self-consistency) equations in powers of order parameter \cite{gor1959microscopic}.
Below we present a short version.
First, we derive potential terms and then kinetic (gradient) terms.
We start with introducing notations $f_s(\bk) = \cos k_x +\cos k_y$, $f_d(\bk) = \cos k_x -\cos k_y$.
Microscopic self-consistency equation \eqref{eq:self consistency 1 component} takes form
\begin{equation}
\begin{gathered}
 \Delta(\bk) = \frac{1}{N} \sum_{\bk'} V \left[ f_s(\bk)f_s(\bk') + f_d(\bk)f_d(\bk') \right] \Delta(\bk') \\
 \cdot T \sum_{\omega_n}\frac{1}{\omega_n^2 + \xi^2(\bk')+|\Delta(\bk')|^2} \\
 \approx \frac{VT}{N} \sum_{\bk'} \left[ f_s(\bk)f_s(\bk') + f_d(\bk)f_d(\bk') \right] \\
 \cdot \sum_{\omega_n} \left[ \frac{\Delta(\bk')}{\omega_n^2 + \xi^2(\bk')} - \frac{\Delta(\bk') |\Delta(\bk')|^2}{[\omega_n^2 + \xi^2(\bk')]^2} \right],
\end{gathered}
\end{equation}
where we used identity $T \sum_{\omega_n} (\omega_n^2+E^2)^{-1} = \frac{\tanh (E/2T)}{2E}$, $\omega_n = (1+2n)\pi T$ are Matsubara frequencies.
Next, equations for $s$ and $d$-wave components of the order parameter $\Delta(\bk) = \Delta_s f_s(\bk) + \Delta_d f_d(\bk)$ can be separated:
\begin{widetext}
    \begin{equation} \label{eq:microscopically derived GL equation}
    \begin{gathered}
        \Delta_s \left(\frac{1}{V} - \frac{T}{N} \sum_{\bk',\omega_n} \frac{f_s^2(\bk')}{\omega_n^2 + \xi^2(\bk')} \right) - \Delta_d \frac{T}{N} \sum_{\bk',\omega_n} \frac{f_s(\bk')f_d(\bk')}{\omega_n^2 + \xi^2(\bk')}
        + \frac{T}{N} \sum_{\bk',\omega_n} \frac{1}{[\omega_n^2 + \xi^2(\bk')]^2}  \\
        \cdot \left[ f_s^4(\bk') |\Delta_s|^2 \Delta_s + 2 f_d^2(\bk') f_s^2(\bk') |\Delta_d|^2 \Delta_s + f_d^2(\bk') f_s^2(\bk') \Delta_d^2 \Delta_s^* \right. \\
        \left.  + f_d^3(\bk') f_s(\bk') |\Delta_d|^2 \Delta_d + f_d(\bk') f_s^3(\bk') \left(2 |\Delta_s|^2 \Delta_d + \Delta_s^2 \Delta_d^* \right) \right] = 0,
    \end{gathered}
    \end{equation}
\end{widetext}
the equation for $\Delta_d$ can be obtained by changing indices $s \leftrightarrow d$.
The sum over $\bk'$ is zero for odd in $f_{s/d}$ terms for the tetragonal system because they are odd w.r.t. $\pi/2$ rotation.
Comparing terms from \eqref{eq:microscopically derived GL equation} and GL equations that follow from \eqref{eq:GL free energy original} one finds
\begin{equation}
\begin{gathered} \label{eq:potential energy coefs}
    \alpha_1 = \frac{1}{V} - \frac{T}{N} \sum_{\bk,\omega_n} \frac{f_s^2(\bk)}{\omega_n^2 + \xi^2(\bk)}, \\
    \alpha_2 = \frac{1}{V} - \frac{T}{N} \sum_{\bk,\omega_n} \frac{f_d^2(\bk)}{\omega_n^2 + \xi^2(\bk)}, \\
    \{\beta_1,\beta_2,\beta_3,\beta_4\} \\
    = \frac{T}{2N} \sum_{\bk,\omega_n} \frac{\{f_s^4(\bk),f_d^4(\bk),4f_s^2(\bk) f_d^2(\bk),f_s^2(\bk) f_d^2(\bk)\}}{[\omega_n^2 + \xi^2(\bk)]^2}.
\end{gathered}
\end{equation}
Note, that $\beta_4 = \beta_3/4.$

Coefficients $\beta_{1-4}$ differ by factor 4 from corresponding coefficients in Refs.~\cite{kuboki2012microscopic, feder1997microscopic}.
Origin of the difference roots in the gap definition: They use $\Delta(\bk) = \Delta_s \frac{f_s(\bk)}{2} + \Delta_d \frac{f_d(\bk)}{2}$.
Therefore, if quadratic in $\Delta$'s terms coincide (situation that we have), quartic terms coefficients should be multiplied by factor of $2^2$ when transforming from $\Delta(\bk) = \Delta_s f_s(\bk) + \Delta_d f_d(\bk)$ to $\Delta(\bk) = \Delta_s \frac{f_s(\bk)}{2} + \Delta_d \frac{f_d(\bk)}{2}$.

Now, let us compute gradient terms.
Below we derive coefficients only for quadratic in gaps $\Delta_{s/d}$ terms.
Self-consistency equation can be written in real space using Green functions $\mathcal{G}_0$ in the absence of a magnetic field ($\boldsymbol{A}=0$):
\begin{equation} \label{eq:self-consistency using Green functions}
 \frac{1}{V} \Delta^*_\eta(\br) = T\sum_{\br',\epsilon,\omega_n} \mathcal{G}_0 (\br',\br + \eta,-\omega_n) \Delta_\epsilon^* (\br') \mathcal{G}_0 (\br'+\epsilon,\br,\omega_n),
\end{equation}
where $\eta, \, \epsilon=\pm \hat{x}, \, \pm \hat{y}$, and Green function is assumed to be translationally invariant
$\mathcal{G}_0 (\br',\br,\omega_n) = \mathcal{G}_0 (\br'-\br,\omega_n) = \frac{1}{N} \sum_{\bk} e^{ i \bk ( \br'-\br )} / (i \omega_n - \xi_{\bk} )$.
Superconducting gap that lives on a links can be expanded in powers of derivatives:
\begin{equation}
\begin{gathered} \label{eq:gap series expansion}
    \Delta_\epsilon (\br') = \Delta_\epsilon(\br) + (\br'-\br)_\mu \nabla_\mu \Delta_\epsilon(\br) \\
    + \frac{1}{2} (\br'-\br)_\mu (\br'-\br)_\nu \nabla_\mu \nabla_\nu \Delta_\epsilon(\br).
\end{gathered}
\end{equation}
Here assumed summation over repeated indices.
Substituting Green functions and \eqref{eq:gap series expansion} to \eqref{eq:self-consistency using Green functions} one gets
\begin{widetext}
    \begin{equation} \label{eq:kinetic part 1}
    \begin{gathered}
        \frac{1}{V} \Delta^*_\eta(\br) = \frac{T}{N^2} \sum_{\bk_1,\bk_2,\epsilon,\omega_n} \frac{1}{-i \omega_n - \xi_{\bk_1}} \frac{1}{i \omega_n - \xi_{\bk_2}} \left[ \Delta^*_\epsilon(\br) \underbrace{\sum_{\br'} e^{i (\bk_1 + \bk_2)(\br'-\br)}}_{N \delta_{\bk_1,\bk_2}}e^{-i \bk_1 \eta + i \bk_2 \epsilon} \right. \\
        \left. +\nabla_\mu \Delta^*_\epsilon(\br) \sum_{\br'} \underbrace{ (\br'-\br)_\mu }_{(-i \frac{\partial}{\partial k_{1\mu}}+\eta_\mu)} e^{i (\bk_1 + \bk_2)(\br'-\br)-i \bk_1 \eta + i \bk_2 \epsilon}
        +\frac{1}{2} \nabla_\mu \nabla_\nu \Delta^*_\epsilon(\br) \sum_{\br'} \underbrace{ (\br'-\br)_\mu }_{(-i \frac{\partial}{\partial k_{1\mu}}+\eta_\mu)} \underbrace{ (\br'-\br)_\nu }_{(-i \frac{\partial}{\partial k_{2\nu}}-\epsilon_\nu)} e^{i (\bk_1 + \bk_2)(\br'-\br)-i \bk_1 \eta + i \bk_2 \epsilon} \right].
    \end{gathered}
    \end{equation}
\end{widetext}
Next, we "integrate by parts" terms containing $\frac{\partial}{\partial k_{1\mu}}$ and $\frac{\partial}{\partial k_{2\nu}}$ and then perform sum over $\br'$ like in the first line of \eqref{eq:kinetic part 1}
\begin{widetext}
    \begin{equation}
    \begin{gathered}
        \frac{1}{V} \Delta^*_\eta(\br) = \frac{T}{N} \sum_{\bk,\epsilon,\omega_n} \frac{1}{\omega_n^2 + \xi_{\bk}^2}  \left[ \Delta^*_\epsilon(\br) + \eta_\mu \nabla_\mu \Delta^*_\epsilon(\br) - \frac{1}{2} \eta_\mu \epsilon_\nu \nabla_\mu \nabla_\nu \Delta^*_\epsilon(\br) + i \nabla_\mu \Delta^*_\epsilon(\br) \frac{\partial \xi_{\bk}/\partial k_{\mu}}{-i \omega_n - \xi_{\bk}}  \right. \\
        \left. - \frac{i}{2} \eta_\mu \nabla_\mu \nabla_\nu \Delta^*_\epsilon(\br) \frac{\partial \xi_{\bk}/\partial k_{\nu}}{i \omega_n - \xi_{\bk}} - \frac{i}{2} \epsilon_\nu \nabla_\mu \nabla_\nu \Delta^*_\epsilon(\br)\frac{\partial \xi_{\bk}/\partial k_{\mu}}{-i \omega_n - \xi_{\bk}} + \frac{1}{2} \nabla_\mu \nabla_\nu \Delta^*_\epsilon(\br) \frac{(\partial \xi_{\bk}/\partial k_{\mu})(\partial \xi_{\bk}/\partial k_{\nu})}{\omega_n^2 + \xi^2_{\bk}} \right]  e^{-i \bk (\eta + \epsilon)}.
    \end{gathered}
    \end{equation}
\end{widetext}
All the terms except the first and the last vanish during summation over $\epsilon$ and $\bk$ because they are odd w.r.t. space inversion (containing $\epsilon, \, \eta$) or $\pi/2$ rotation (with $\partial \xi_{\bk}/\partial k_{\mu}$).
The last term is nonzero only for $\mu=\nu$ for the same reason.
These facts significantly simplify self-consistency equation (written up to linear order)
\begin{equation} \label{eq:kinetic part 2}
    \begin{gathered}
        \frac{1}{V} \Delta^*_\eta(\br) = \frac{T}{N} \sum_{\bk,\epsilon,\omega_n} \frac{1}{\omega_n^2 + \xi_{\bk}^2}  \left[ \Delta^*_\epsilon(\br) \right. \\
        \left. + \frac{1}{2} \frac{\partial^2}{\partial \br_\mu^2} \Delta^*_\epsilon(\br) \frac{(\partial \xi_{\bk}/\partial k_{\mu})^2}{\omega_n^2 + \xi^2_{\bk}} \right]  e^{-i \bk (\eta + \epsilon)}.
    \end{gathered}
\end{equation}
Final step is to introduce extended $s$-wave and $d$-wave order parameter [$\Delta_s = \frac{1}{2} (\Delta_{+\hat{x}} + \Delta_{-\hat{x}} + \Delta_{+\hat{y}} + \Delta_{-\hat{y}})$, $\Delta_d = \frac{1}{2} (\Delta_{+\hat{x}} + \Delta_{-\hat{x}} - \Delta_{+\hat{y}} - \Delta_{-\hat{y}})$], write self-consistency \eqref{eq:kinetic part 2} for these irreducible representations and compute sum over $\epsilon$:
\begin{equation}
\begin{gathered}
    \Delta^*_s \left(\frac{1}{V} - \frac{T}{N} \sum_{\bk,\omega_n} \frac{f_s^2(\bk)}{\omega_n^2 + \xi^2(\bk)} \right) \\
    - 2 \frac{T}{N} \sum_{\bk,\omega_n} \frac{f_s(\bk) [\sin^2 (k_x) \frac{\partial^2}{\partial x^2} + \sin^2 (k_y) \frac{\partial^2}{\partial y^2}]}{[\omega_n^2 + \xi^2(\bk)]^2} \\
    \cdot [\Delta^*_{s} f_s(\bk) + \Delta^*_{d} f_d(\bk) ] = 0.
\end{gathered}
\end{equation}
Here we kept only terms that do not vanish after summation over $\bk$.
Comparing this equation to linearized GL equations from \eqref{eq:GL free energy original},
gradient coefficients take form
\begin{equation}
\begin{gathered} \label{eq:gradient terms}
    \{\gamma_1,\gamma_2,\gamma_{12}\} \\
    = 2\frac{T}{N} \sum_{\bk,\omega_n} \sin^2 (k_x) \frac{\{ f_s^2(\bk), f_d^2(\bk), - f_s(\bk) f_d(\bk) \}}{[\omega_n^2 + \xi^2(\bk)]^2},
\end{gathered}
\end{equation}
where $\gamma_2$ is derived by analogy.

Inclusion of gauge filed $\boldsymbol{A}$ changes derivative $\partial_i$ to covariant one $D_i=\partial_i + i\frac{2e a}{\hbar} A_i$, where factor of 2 emerges due to two electron coupling in the gap $\Delta$.

The free energy density \eqref{eq:GL free energy original} is defined up to a multiplication factor
which can be found by comparison of energies in GL model and microscopic model at $T=0$.
In the Bardeen-Cooper-Schrieffer (BCS) limit it is density of states on the Fermi level.

In the BCS limit ($T \ll \omega_D \ll 1$) GL coefficients can be simplified
\begin{equation} \label{eq:alpha BCS limit}
\begin{gathered}
    \alpha_1 \approx \frac{1}{V} - 2 \langle N_F f_s^2(\bk) \rangle_\text{FS} \int_0^{\omega_D} d \xi \frac{\tanh{\frac{\xi}{2T}}}{2 \xi} \\
    \approx \langle N_F f_s^2(\bk) \rangle_\text{FS} \ln \frac{T}{T_s},
\end{gathered}
\end{equation}
where $T_s$ is defined in \eqref{eq:s wave critical temperature}.
Similarly $\alpha_2 \approx \langle N_F f_d^2(\bk) \rangle_\text{FS} \ln \frac{T}{T_d}$.
Coefficients $\beta_{1-4}$ and $\gamma$'s have similar structure and can be simplified in the following way
\begin{equation} \label{eq:beta BCS limit}
\begin{gathered}
    \beta_1 \approx \langle N_F f_s^4(\bk) \rangle_\text{FS} \int_0^{\omega_D} \frac{d \xi}{2 \xi^2} \left( \frac{\tanh{\frac{\xi}{2T}}}{2 \xi} - \frac{1}{4T \cosh^2 \frac{\xi}{2T}}  \right) \\
    \approx \langle N_F f_s^4(\bk)\rangle_\text{FS} \frac{7 \zeta(3)}{16 \pi^2 T^2}.
\end{gathered}
\end{equation}
As a special case, one can obtain GL coefficients for local $s$-wave pairing derived by Gor’kov \cite{gor1959microscopic}.
Considering $f_s = 1$, $f_d = 0$ we recover potential terms from Ref.~\cite{gor1959microscopic}.

\section{Details of Ginzburg--Landau calculations}

GL parameters for points in \figref{fig:s+id phase diagram wD=0.1} were calculated using Eqs.~(\ref{eq:potential energy coefs}), (\ref{eq:gradient terms}) within BCS approximation [like Eqs.~(\ref{eq:alpha BCS limit}), (\ref{eq:beta BCS limit})].
Averaging of form factors over Fermi surface is performed numerically.
They are presented in \tableref{tab:model parameters}.
Recall that coefficient $\beta_4 = \beta_3/4$.

\begin{table}
    \centering
    \begin{tabular}{l | c|c|c | l}
         &  Set 1&  Set 2&  Set 3& Units\\ \hline
         Figure&  3(a), 4, 6&  3(b), 7&  3(c)& \\ 
         Ground state&  $s+id$&  $s$-wave&  $d$-wave& \\ \hline
         $\mu$ & 1.665&  1.673&  1.665& $t_{xy}$\\
         $T$ &  2.1 &  2.48 &  2.4& $\times 10^{-4}t_{xy}$  \\ \hline
         $\alpha_1$ & $-$13.07&  $-$3.369&  $-$2.185& $\times 10^{-3}$ \\
         $\alpha_2$ & $-$16.28&  1.872&  $-$5.321& $\times 10^{-3}$ \\
         $\beta_1$ & 6.829&  4.982&  5.229& $\times 10^4$  \\
         $\beta_2$ & 10.22&  7.209&  7.821& $\times 10^4$ \\
         $\beta_3$ & 2.749&  1.972&  2.104&  $\times 10^5$ \\
         $\gamma_1$ & 2.571&  1.857&  1.969&  $\times 10^5$\\
         $\gamma_2$ & 2.257&  1.604&  1.728&  $\times 10^5$\\
         $\gamma_{12}$ & 1.374&  0.9860&  1.052&  $\times 10^5$
    \end{tabular}
 \caption{Microscopically derived Ginzburg--Landau free energy parameters for energy cutoff $\omega_D=0.1$ for Eq.~(\ref{eq:GL free energy original}).}
 \label{tab:model parameters}
\end{table}

It is convenient to rescale space (and simultaneously all GL coefficients) for numerical calculations. Total free energy in two dimensions has form
\begin{equation}
    \mathcal{F} = \int_{\mathcal{R}^2} F(x) d^2 x,
\end{equation}
where $F(x)$ is given by \eqref{eq:GL free energy original}.
Performing space $x = \sqrt{\frac{\gamma_1}{|\alpha_1|}} x'$ and field $\Delta_{s/d}(x)= \sqrt{\frac{|\alpha_1|}{\beta_1}}\tilde{\Delta}_{s/d}(x')$, $\boldsymbol{\tilde{A}}(x) = \frac{1}{q} \sqrt{\frac{|\alpha_1|}{\gamma_1}} \tilde{\boldsymbol{A}}'(x')$ rescaling one gets
\begin{equation}
\begin{gathered}
    \mathcal{F} = \frac{|\alpha_1| \gamma_1}{\beta_1} \int_{\mathcal{R}^2} d^2 x' \left[ \frac{\alpha_1}{|\alpha_1|} |\tilde{\Delta}_s|^2 + \frac{\alpha_2}{|\alpha_1|} |\tilde{\Delta}_d|^2 \right. \\
     + |\tilde{\Delta}_s|^4 + \frac{\beta_2}{\beta_1} |\tilde{\Delta}_d|^4 + \frac{\beta_3}{\beta_1} |\tilde{\Delta}_s|^2 |\tilde{\Delta}_d|^2 \\
     + \frac{\beta_4}{\beta_1} (\tilde{\Delta}_s^{*2} \tilde{\Delta}_d^2 + \tilde{\Delta}_s^2 \tilde{\Delta}_d^{*2}) + |\boldsymbol{D}' \tilde{\Delta}_s|^2 + \frac{\gamma_2}{\gamma_1} |\boldsymbol{D}' \tilde{\Delta}_d|^2 \\
    + \frac{\gamma_{12}}{\gamma_1}\left[ (D'_y \tilde{\Delta}_s)^* (D'_y \tilde{\Delta}_d) - (D'_x \tilde{\Delta}_s)^* (D'_x \tilde{\Delta}_d) + \text{c.c.} \right] \\
    + \frac{\kappa^2}{2} (\nabla' \times \tilde{\boldsymbol{A}}')^2,
\end{gathered}
\end{equation}
where $D'_i = \partial'_i + \tilde{A}_i'$ is a rescaled covariant derivative,
$\kappa = \frac{\sqrt{\beta_1}}{\gamma_1 q}$.
This form of free energy allows one to have all GL coefficients to be $O(1)$ that is convenient for numerical methods.

\subsection{Coherence lengths calculation} \label{app:coherence length}

Exact evaluation leads to a degree 6 polynomial equation (for amplitudes, phase difference, and three gauge field components) that determine all the lengths scales in the system, while all the normal modes are in general mixed \cite{speight2021magnetic}.
Here we provide rough approximations to the coherence lengths corresponding to amplitude variations and relative phase variations.
This is done by treating both cases separately with no mixing: Phase difference is fixed for calculations of coherence lengths for amplitudes and vice versa.
Second simplification we make: Translation invariance orthogonal to perturbation direction.
Both these simplifications in fact break for vortex perturbation.
However, this method allows us to get first coarse estimate for coherence lengths to search for type-1.5 superconductivity regime.

Let us linearize $F$ with respect to variations in the order parameter amplitudes ($|\Delta_{s/d}| = |\Delta_{s/d}^\text{GS}| + \sigma_{s/d}$ where $\sigma_{s/d}$ is small).
Assuming amplitude variation in the direction $\boldsymbol{n}=(\cos \varphi, \sin \varphi)$ free energy density becomes
\begin{equation}
\begin{gathered}
    F_\text{lin} = F_\text{GS} + \frac{1}{2} (\nabla \sigma_\alpha) \Gamma_{\alpha \beta} (\nabla \sigma_\beta) + \frac{1}{2} \sigma_\alpha^* \mathcal{H}_{\alpha \beta} \sigma_\beta
\end{gathered}
\end{equation}
where $\Gamma(\varphi)$ is given in \eqref{eq:gamma matrix}, the Hessian $\mathcal{H}$ is
\begin{equation}
    \mathcal{H} =
    \begin{pmatrix}
        8 \beta_1 |\Delta_{s}^\text{GS}|^2 & 4 \tilde\beta_3(\theta^\text{GS}) |\Delta_{s}^\text{GS}| |\Delta_{d}^\text{GS}| \\
        4 \tilde\beta_3(\theta^\text{GS}) |\Delta_{s}^\text{GS}| |\Delta_{d}^\text{GS}| & 8 \beta_2 |\Delta_{d}^\text{GS}|^2
    \end{pmatrix}
\end{equation}
with $\tilde{\beta}_3(\theta)= \beta_3 + 2\beta_4\cos2\theta$ and $\theta^\text{GS}=\pi/2$.

Coherence lengths for amplitudes $\xi_i$ within the approximation are given by
\begin{equation}
    \xi_i(\bn) = \frac{a}{\sqrt{\eta_i(\varphi)}},
\end{equation}
where $\eta_i(\varphi)$ are eigenvalues of $\Gamma^{-1}(\varphi) \mathcal{H}$.
Coherence lengths $\xi_i(\bn)$ are associated with linear combinations of $\sigma_s$ and $\sigma_d$ that is called hybridization.
Moreover, presence of off-diagonal terms in $\Gamma(\varphi)$ ensures that system has two coherence lengths for amplitudes even for ground state with pure $s$-wave or pure $d$-wave.

Let us now consider the relative phase mode and ignore any mixed gradient couplings for simplicity.
That is, we consider phase-difference-only fluctuations.
Let us consider the fluctuation of the relative phase about the ground state,
\begin{equation}
    \theta = \theta^\text{GS} + \phi,
\end{equation}
where the small fluctuation is $\phi$.
The linearized free energy with respect to $\phi$ is
\begin{equation}
\begin{gathered}
    F_{\text{lin } \phi} = F_\text{GS} + \frac{1}{2} K_{\textup{rel}}(\varphi,\theta^\text{GS}) (\nabla \phi)^2 + \frac{1}{2} \left.\frac{\partial^2 F_p}{\partial \theta^2}\right|_{\theta=\theta^\text{GS}} \phi^2,
\end{gathered}
\end{equation}
where $F_p$ is potential part of the free energy \eqref{eq:GL free energy original},
\begin{widetext}
    \begin{equation}
        K_{\textup{rel}}(\varphi,\theta) 
        = \frac{2|\Delta_{s}^\text{GS}|^2 |\Delta_{d}^\text{GS}|^2 (\gamma_1 \gamma_2 - \gamma_{12}^2\cos^22\varphi \cos^2\theta)}{\gamma_1 |\Delta_{s}^\text{GS}|^2 - 2\gamma_{12} |\Delta_{s}^\text{GS}| |\Delta_{d}^\text{GS}| \cos2\varphi\cos\theta+\gamma_2|\Delta_{d}^\text{GS}|^2}.
    \end{equation}
\end{widetext}
When substituting $\theta=\theta^\text{GS}$, one gets
\begin{equation}
    K_{\textup{rel}}(\varphi,\theta^\text{GS}) = \frac{2\gamma_1 \gamma_2 |\Delta_{s}^\text{GS}|^2 |\Delta_{d}^\text{GS}|^2}{\gamma_1|\Delta_{s}^\text{GS}|^2 + \gamma_2 |\Delta_{d}^\text{GS}|^2}
\end{equation}
and
\begin{equation}
    \left.\frac{\partial^2 F_p}{\partial \theta^2}\right|_{\theta=\theta^\text{GS}}=2\beta_3 |\Delta_{s}^\text{GS}|^2 |\Delta_{d}^\text{GS}|^2.
\end{equation}
Therefore coherence length for relative phase variations
\begin{equation}
    \xi_\theta(\varphi) = a \sqrt{\frac{\gamma_1 \gamma_2}{\beta_3(\gamma_1|\Delta_{s}^\text{GS}|^2 + \gamma_2 |\Delta_{d}^\text{GS}|^2)}}
\end{equation}
is independent of the direction $\varphi$.
Note that it is defined in the region with two nucleated gap components ($s+id$ phase in our case) as can be seen in \figref{fig:xi and lambda}.

\subsection{Numerical calculation details}

After making rough estimates for length scales we can move to numerical solutions that allow to calculate length scales without these simplified estimates.
The model consists of a two-component complex order parameter $\Delta_\alpha\in\mathbb{C}$ with $\alpha=1,2$ and a vector gauge field $\vec{\tilde{A}}=(\tilde{A}_1, \tilde{A}_2)\in\mathbb{R}^2$.
We identify $\Delta_s=\Delta_1$ and $\Delta_d=\Delta_2$.
Let the local coordinate be $x=(x_1,x_2)\in\mathbb{R}^2$ and define the local frame $\partial_j=\partial/\partial x_j$ for $j=1,2$.
The GL energy density functional in index notation is
\begin{equation}
\label{eq: GL energy}
    F =  \frac{1}{2} \gamma_{jk}^{\alpha\beta} \overline{D_j \Delta_\alpha} D_k \Delta_\beta + \frac{\kappa^2}{2}|\vec{\nabla}\times\vec{\tilde{A}}|^2 + F_{p}(\Delta_s,\Delta_d),
\end{equation}
where $\alpha,\beta,j,k=1,2$ and the gauge covariant derivative is defined by the vector $\vec{D}=\vec{\nabla}+i\vec{\tilde{A}}$, and $\vec{\nabla}=(\partial_1,\partial_2)$ is the gradient operator.
Note that we define
\begin{equation}
    \overline{D_j\Delta_\alpha} = \partial_j \bar{\Delta}_\alpha -i\tilde{A}_j\bar{\Delta}_\alpha.
\end{equation}
The anisotropy matrices are given by
\begin{equation}
    \gamma_{11} = 2
    \begin{pmatrix}
        \gamma_1 & -\gamma_{12} \\
        -\gamma_{12} & \gamma_2
    \end{pmatrix},
    \quad
    \gamma_{22} = 2
    \begin{pmatrix}
        \gamma_1 & \gamma_{12} \\
        \gamma_{12} & \gamma_2
    \end{pmatrix}.
\end{equation}
The potential energy we are considering is defined by
\begin{align}
    &F_{p}(\Delta_s,\Delta_d) = \alpha_1 |\Delta_s|^2 + \alpha_2 |\Delta_d|^2 + \beta_1 |\Delta_s|^4 + \beta_2 |\Delta_d|^4  \nonumber \\
    & + \beta_3 |\Delta_s|^2 |\Delta_d|^2+ \beta_4 \left( \Delta_s^2 \bar{\Delta}_d^2 + \bar{\Delta}_s^2 \Delta_d^2 \right).
\end{align}
The associated static GL field equations are found to be
\begin{subequations}
\label{eq: GL field equations}
    \begin{align}
        \frac{\delta F}{\delta \bar{\Delta}_\alpha} = \, & \frac{\partial F_p}{\partial \bar{\Delta}_\alpha} - \frac{1}{2}\gamma_{ij}^{\alpha\beta}D_i D_j \Delta_\beta, \\
        \frac{\delta F}{\delta \tilde{A}_i} = \, & J_i + \kappa^2 \partial_i \partial_j \tilde{A}_j - \kappa^2\partial_j \partial_j \tilde{A}_i.
    \end{align}
\end{subequations}
where the supercurrent is
\begin{equation}
    J_i = \frac{i}{2} \left( \gamma_{ji}^{\alpha\beta} \Delta_\beta \overline{D_j\Delta_\alpha} - \gamma_{ij}^{\beta\alpha}\bar{\Delta}_\beta D_j \Delta_\alpha \right).
\end{equation}

To obtain vortex solutions in our model, we must minimize the GL free energy \eqref{eq: GL energy}.
This amounts to solving the corresponding field equations \eqref{eq: GL field equations}.
We do this using an accelerated gradient descent method with flow arresting criteria.
We reformulate the minimization as a second order dynamical problem and solve the second order system
\begin{align}
    \frac{\textup{d}^2\Delta_\alpha}{\textup{d}t^2} = \, & \frac{1}{2}\gamma_{ij}^{\alpha\beta}D_i D_j \Delta_\beta - \frac{\partial F_p}{\partial \bar{\Delta}_\alpha}, \\
    \frac{\textup{d}^2 \tilde{A}_i}{\textup{d}t^2} = \, & \kappa^2 \partial_j (\partial_j \tilde{A}_i - \partial_i \tilde{A}_j) - J_i,
\end{align}
where $t$ is a fictitious time.
This system can then be reduced to a coupled first order system, which we solve using a fourth order Runge-Kutta method.

In order for our method to converge to a $N$-vortex solution, we need some appropriate initial configuration $\{\Delta_{s/d}(0),\vec{\tilde{A}}(0)\}=\{\Delta_{s/d}^0,\vec{A}^0\}$.
For the superconducting order parameter, we use the ansatz
\begin{equation}
\label{eq: Nielsen-Olesen ansatz}
    \begin{split}
        \Delta_\alpha^0 = \Delta_{\alpha}^\text{GS} \sigma_\alpha(r)e^{iN_\alpha\theta_\alpha}, \quad \vec{A}^0 = \frac{Na(r)}{r} \left(-\sin\theta,\cos\theta\right).
    \end{split}
\end{equation}
where $N_\alpha$ is the vortex number in each component, $N$ is the total vortex number and $|\Delta_{\alpha}^\text{GS}|$ is the ground state value of the associated component.
The total magnetic flux through the plane is given by
\begin{equation}
    \Phi = {2\pi}N, \quad N = \frac{|\Delta_{s}^\text{GS}|^2 N_s + |\Delta_{d}^\text{GS}|^2 N_d}{|\Delta_{s}^\text{GS}|^2 + |\Delta_{d}^\text{GS}|^2}.
\end{equation}
The profile functions are monotonically increasing functions that satisfy the boundary conditions $\sigma_\alpha(0)=a(0)=0$ and 
\begin{equation}
    \lim_{r\rightarrow\infty}\sigma_\alpha(r)=\lim_{r\rightarrow\infty}a(r)=1.
\end{equation}
Note that in general boundary conditions are more subtle \cite{samoilenka2021microscopic}.
However, we choose simulation box large enough so that there is no vortex interaction with boundaries 
within numerical accuracy.
Hence the simplest boundary conditions are sufficient for our case.

For our initial configuration, we simply set $\sigma_\alpha(r)=a(r)=\tanh(r)$.

\subsection{Skyrmions}

When integer flux vortices split into spatially separated fractional vortices in each component $\Delta_\alpha$, such that their cores (zeroes of the components $\Delta_\alpha=0$) are never coincident, skyrmions form.
That can be seen by representing superconducting gap fields in terms of a pseudospin vector \cite{Babaev.Faddeev.ea:02}.
For such structures, one may construct a gauge invariant field $\phi:\mathbb{R}^2\rightarrow S^2$, given by
\begin{equation}
    \phi= \frac{1}{|\Delta_s|^2+|\Delta_d|^2} 
    \begin{pmatrix}
        \Delta_s^*\Delta_d + \Delta_s\Delta_d^* \\
        i(\Delta_s^*\Delta_d - \Delta_s\Delta_d^*) \\
        |\Delta_d|^2-|\Delta_s|^2        
    \end{pmatrix}.
\end{equation}
We can describe the skyrmion field using the gauge invariant quantities $\Delta$, $\rho=\sqrt{|\Delta_s|^2+|\Delta_d|^2}$, and the supercurrent $J$, where we are regarding the supercurrent as a 1-form on $\mathbb{R}^2$.
We can express the magnetic field (as a 2-form on $\mathbb{R}^2$) in terms of these gauge invariant quantities as \cite{winyard2019skyrmion,Garaud.Carlstrom.ea:13}
\begin{equation}
    B = \frac{1}{2}\phi^*\omega - \textup{d}\left( \frac{J}{\rho^2} \right),
\end{equation}
where $\omega$ is the usual area form on the target $S^2$ and $\textup{d}$ is the exterior derivative.
Then, using Stoke's Theorem, we observe that the skyrmion number is precisely the number of magnetic flux quanta in the field configuration, that is
\begin{equation}
    \int_{\mathbb{R}^2} B = \frac{1}{2}\int_{\mathbb{R}^2} \phi^*\omega = 2\pi \mathcal{Q}(\phi).
\end{equation}
The skyrmion number $\mathcal{Q}$ can be computed using the following integral representation, see a detailed discussion in \cite{Garaud.Carlstrom.ea:13}
\begin{equation}
    \mathcal{Q}(\phi) = \frac{i\epsilon_{ji}}{2\pi}\int_{\mathbb{R}^2} \textup{d}^2x \, \frac{1}{|\Psi|^4} \left( |\Psi|^2 \partial_i\Psi^\dagger \partial_j\Psi + \Psi^\dagger \partial_i \Psi \partial_j \Psi^\dagger \Psi \right),
\end{equation}
where $\Psi=(\Delta_s,\Delta_d)$.

\subsection{Additional data} \label{app:additional data}

\begin{figure*}[t]
    \centering
    \includegraphics[width=0.95\textwidth]{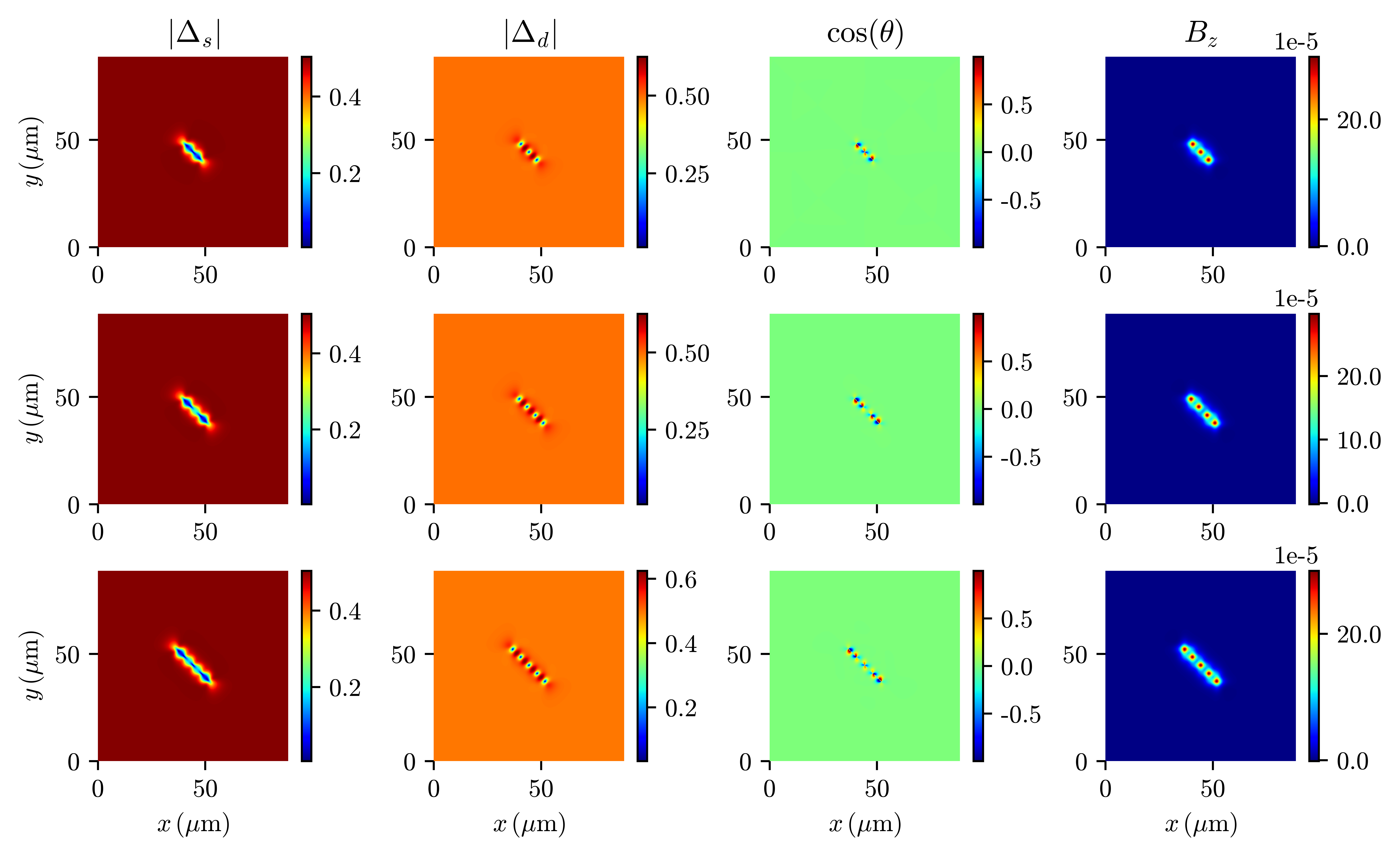}
    \caption{Vortex cluster solutions for $s+id$ ground state in Ginzburg--Landau model.
    Number of flux quanta: (a) 3, (b) 4, (c) 5.
    Model parameters correspond to green point in \figref{fig:s+id phase diagram wD=0.1}, it is the first column in \tableref{tab:model parameters}.
    Panels 1-4 show amplitudes $|\Delta_s|$, $|\Delta_d|$, relative phase difference $\theta=\arg (\Delta_s \Delta_d^*)$, and magnetic field $B_z$.
    Vortex interaction energy is negative that can be seen in \figref{fig: Binding energies}(a).
    The total energy density qualitatively follows the magnetic field shown in panels 4.
    Boundary conditions correspond to no current flowing through the
boundary, external magnetic field is zero.
    Gap amplitudes are presented in units $t_{xy} \sqrt{|\alpha_1|/\beta_1}$, magnetic field $B_z$ in $2 \hbar |\alpha_1|/(e a^2 \gamma_1)$.}
    \label{fig: type-1.5 s+id state}
\end{figure*}

\begin{figure*}[t]
    \centering
    \includegraphics[width=0.95\textwidth]{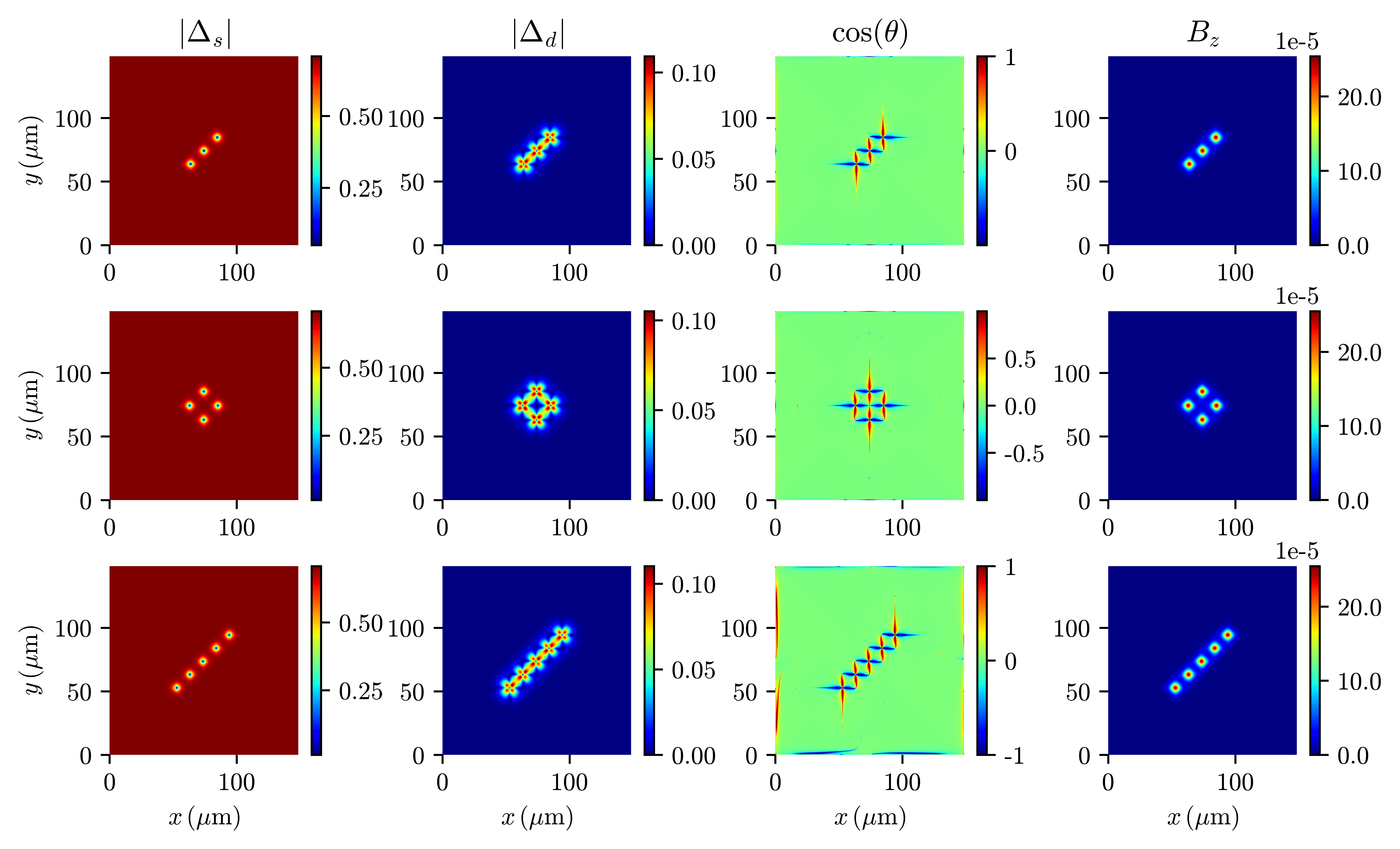}
    \caption{Vortex cluster solutions for pure $s$-wave ground state in Ginzburg--Landau model.
    Number of flux quanta: (a) 3, (b) 4, (c) 5.
    Model parameters correspond to black point in \figref{fig:s+id phase diagram wD=0.1}, it is the second column in \tableref{tab:model parameters}.
    Panels 1-4 show amplitudes $|\Delta_s|$, $|\Delta_d|$, relative phase difference $\theta=\arg (\Delta_s \Delta_d^*)$, and magnetic field $B_z$.
    Vortex interaction energy is negative, that can be seen in \figref{fig: Binding energies}(b).
    The total energy density qualitatively follows the magnetic field shown in panels 4.
    Boundary conditions correspond to no current flowing through the
boundary, external magnetic field is zero, and the sample size is chosen so that vortex interaction with boundaries is negligible.
    Gap amplitudes are presented in units $t_{xy} \sqrt{|\alpha_1|/\beta_1}$, magnetic field $B_z$ in $2 \hbar |\alpha_1|/(e a^2 \gamma_1)$.
    }
    \label{fig: type-1.5 s-wave}
\end{figure*}

In the appendix we provide multivortex cluster solutions for different flux quanta and two ground states: $s+id$ state (\figref{fig: type-1.5 s+id state}) and pure $s$-wave (\figref{fig: type-1.5 s-wave}).
Solutions for $d$-wave ground state are qualitatively similar to the ones in \figref{fig: type-1.5 s-wave}.
This has origin from the symmetry of the model under swapping of the order parameters $\{\Delta_s,\alpha_1,\beta_1,\gamma_1\} \leftrightarrow \{\Delta_d,\alpha_2,\beta_2,\gamma_2\}$.
The similarity of solutions can be observed in Figs.~\ref{fig: type-1.5}(b) and \ref{fig: type-1.5}(c).

\begin{figure*}[t]
    \centering
    \includegraphics[width=0.45\textwidth]{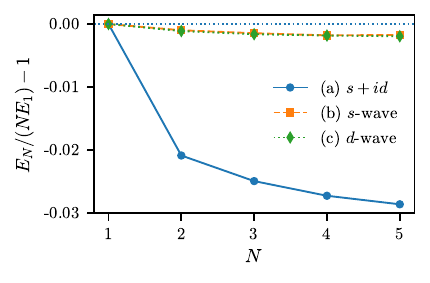}
    \caption{The normalized binding energy in a vortex cluster per vortex $E_{\textup{bind}}/E_1=E_N/(NE_1)-1$ shown for the three distinct regimes.
    Parameter sets correspond to green ($\mu=1.665$, $T=2.1\cdot10^{-4}$), black ($\mu=1.673$, $T=2.48\cdot10^{-4}$) and red ($\mu=1.665$, $T=2.4\cdot10^{-4}$) points in \figref{fig:s+id phase diagram wD=0.1}.
    Ginzburg--Landau model parameters are presented in \tableref{tab:model parameters}.
    }
    \label{fig: Binding energies}
\end{figure*}

Figure~\ref{fig: Binding energies} shows normalized vortex interaction energies.
The binding energy is negative for type-1.5 superconductivity regime.
Note that vortex interaction energy for $s+id$ ground state is one order of magnitude larger than for pure $s$-wave or pure $d$-wave.
Another striking feature is that in contrast to multiband $s$-wave superconductor \cite{nonpairwise}, the vortex clustering in the system with $s$-wave and $d$-wave pairing channels has a tendency to have stripe-like morphology (cf. anisotropic $s$-wave models \cite{winyard2019hierarchies}).

 \bibliography{references}

 \end{document}